\documentclass[12pt,twoside]{report}

\usepackage{thesis}
\usepackage{epsf}
\usepackage{natbib}
\usepackage{graphics}

\newlength{\figwidth} 
\newlength{\appfigwidth} 
\newlength{\dfspace} 

\begin{document}
\pagestyle{empty}
\pagenumbering{roman}

\begin{center}

\vspace*{2.0in}
Copyleft \raisebox{6pt}{\rotatebox{180}{\copyright}} 2001\\
\smallskip\smallskip
by \\
\smallskip\smallskip
Travis Scott Metcalfe \\
\vspace*{3.5in}

This information is free; you can redistribute it\\
under the terms of the GNU General Public License\\
as published by the Free Software Foundation.

\end{center}

\vfill\cleardoublepage\thispagestyle{empty}

\begin{center}
\vspace*{-0.5in}
The Dissertation Committee for Travis Scott Metcalfe\\
Certifies that this is the approved version of the following dissertation:
\bigskip
\vspace*{0.175in}
{\Large \bf COMPUTATIONAL ASTEROSEISMOLOGY}\\
\bigskip
\end{center}
\begin{flushleft}
\vspace*{2.0in}
\hskip 2.5in
{\large COMMITTEE:}\\
\vspace*{0.6in}
\hskip 2.5in
{\underline{~~~~~~~~~~~~~~~~~~~~~~~~~~~~~~~~~~~~~~~~~~~~~~~~~~~~~}}\\
\hskip 2.5in{\small R. Edward Nather, Co-Supervisor}\\
\vspace*{0.35in}
\hskip 2.5in
{\underline{~~~~~~~~~~~~~~~~~~~~~~~~~~~~~~~~~~~~~~~~~~~~~~~~~~~~~}}\\
\hskip 2.5in{\small Donald E. Winget, Co-Supervisor}\\
\vspace*{0.35in}
\hskip 2.5in
{\underline{~~~~~~~~~~~~~~~~~~~~~~~~~~~~~~~~~~~~~~~~~~~~~~~~~~~~~}}\\
\hskip 2.5in{\small Paul Charbonneau}\\
\vspace*{0.35in}
\hskip 2.5in
{\underline{~~~~~~~~~~~~~~~~~~~~~~~~~~~~~~~~~~~~~~~~~~~~~~~~~~~~~}}\\
\hskip 2.5in{\small Kepler Oliveira}\\
\vspace*{0.35in}
\hskip 2.5in
{\underline{~~~~~~~~~~~~~~~~~~~~~~~~~~~~~~~~~~~~~~~~~~~~~~~~~~~~~}}\\
\hskip 2.5in{\small J. Craig Wheeler}\\
\end{flushleft}

\vfill\cleardoublepage\thispagestyle{empty}

\begin{center}
\vspace*{0.35in}
{\Large \bf COMPUTATIONAL ASTEROSEISMOLOGY}\\
\smallskip\smallskip\smallskip\smallskip\smallskip
\vspace*{0.5in}
{\large by}\\
\vspace*{0.5in}
{\large \bf TRAVIS SCOTT METCALFE, B.S., M.A.}\\
\vspace*{1.5in}
{\normalsize \bf DISSERTATION}\\
\smallskip\smallskip
{\normalsize Presented to the Faculty of the Graduate School of}\\
\smallskip\smallskip
{\normalsize The University of Texas at Austin}\\
\smallskip\smallskip
{\normalsize in Partial Fulfillment}\\
\smallskip\smallskip
{\normalsize of the Requirements}\\
\smallskip\smallskip
{\normalsize for the Degree of}\\
\vspace*{0.25in}
{\normalsize \bf DOCTOR OF PHILOSOPHY}\\
\vspace*{1.25in}
{\normalsize THE UNIVERSITY OF TEXAS AT AUSTIN}\\
\smallskip\smallskip
{\normalsize August~~2001}\\
\end{center}

\vfill\cleardoublepage\thispagestyle{empty}

\begin{center}

\vspace*{2.0in}

\parbox[t]{2.0in}{
We are the stars which sing, \\
we sing with our light;      \\
we are the birds of fire,    \\
we fly over the sky.         \\

{\raggedleft {\it ---Dead Can Dance} \\ 
{\vskip -3pt \small Song of the Stars} \\
}}
\end{center}

\vfill\cleardoublepage\thispagestyle{empty}

\normalsize
\pagestyle{plain}

\setcounter{page}{9}

\addcontentsline{toc}{chapter}{Acknowledgements}

\chapter*{Acknowledgements}

When I was looking for the right graduate program in astronomy during the
spring of 1996, I only visited three places. Texas was the last of the
three, and my visit was only a few days before the universal deadline for
making a decision on where to go. I formed my first impression of Ed
Nather and Don Winget while we ate lunch and talked in the WET lab during
my visit. By the end of our discussion, I knew I would come to Texas and
work with them.

Thanks to the unwavering support of my advisers, I am the first of my
class to finish. Only four of my ten original classmates are still working 
toward their Ph.D.; If not for Ed and Don, I might not have continued in 
the program myself. I feel a great sense of privilege for having learned 
to be an astronomer from them. The world needs more scientists like 
Ed Nather and Don Winget.

In addition to our experiences in the lab, I had the pleasure of helping
Ed in the classroom during his last three years of teaching. If I ever
find myself in a position to teach, I will certainly aspire to be as good
as Ed. By my second year as a teaching assistant for him, Ed said that I
didn't need to attend the lectures anymore, but I continued going because
I was getting as much out of them as the students. Besides, it allowed me
to get out of our prison-like building at least a few times a week. Some
of the best conversations I've had with Ed took place during our walks 
over to the Welch Hall classroom.

While developing the metacomputer described in this dissertation I sought
advice and help from Mark Cornell, Bill Spiesman, and Gary Hansen---who
also arranged for the donation of 32 computer processors from AMD. The
software development owes thanks to Mike Montgomery, Paul Bradley, Matt
Wood, Steve Kawaler, Peter Diener, Peter Hoeflich, and especially Paul
Charbonneau and Barry Knapp for providing me with an unreleased improved
version the PIKAIA genetic algorithm. I thank S.O.~Kepler, Craig Wheeler,
Atsuko Nitta and Scott Kleinman for helpful discussions, and Maurizio
Salaris for providing me with data files of white dwarf internal chemical
profiles.

There are many other people who influenced me in earlier stages of my
career, and my dissertation project owes a debt to all of them. I'd like
to thank Terry Bressi and Andrew Tubbiolo for getting me started with
Linux in 1994. Brad Castalia was the first to suggest PVM to me for
parallel computing, and Joe Lazio introduced me to genetic algorithms. I
thank Ray White, Rob Jedicke, Dave Latham, Jim Cordes, Don McCarthy,
Andrea Ghez, Todd Henry, Ted Bowen and Patrick McGuire for their part in
helping me to become a good scientist.

My interest in astronomy was inspired by Darla Casey and Bonnie Osbourne,
and was nurtured by Brian Montgomery, Rick Letherer, and Ed Fitzpatrick.  
I also owe thanks to my parents and the rest of my family in Oregon, who
always believed in me and never questioned my sometimes impractical
decisions. Finally, I'd like to thank Cherie Goff for helping me to
maintain my sanity when times were tough, and for being the ideal
companion the rest of the time.

I am grateful to the High Altitude Observatory Visiting Scientist Program
for fostering this project in a very productive environment for two months
during the summer of 2000. This work was supported by grant NAG5-9321 from
the Applied Information Systems Research Program of the National
Aeronautics \& Space Administration, and in part by grants AST 98-76730
and AST 93-15461 from the National Science Foundation.

\begin{flushright}
{\it June 2001}
\end{flushright}

\vfill\cleardoublepage

\addcontentsline{toc}{chapter}{Abstract}

\begin{center}
\vspace*{0.375in}
{\Large \bf COMPUTATIONAL ASTEROSEISMOLOGY}\\
\smallskip
\vspace*{0.3in}
Publication No.{\underline{~~~~~~~~~~~~~~~}}\\
\vspace*{0.15in}
Travis Scott Metcalfe, Ph.D.\\
The University of Texas at Austin, 2001\\
\bigskip
Co-Supervisors: R. Edward Nather and Donald E. Winget\\
\end{center}
\vspace*{0.3in}

White dwarf asteroseismology offers the opportunity to probe the structure
and composition of stellar objects governed by relatively simple physical
principles. The observational requirements of asteroseismology have been
addressed by the development of the Whole Earth Telescope, but the
analysis procedures still need to be refined before this technique can
yield the complete physical insight that the data can provide. We have
applied an optimization method utilizing a genetic algorithm to the
problem of fitting white dwarf pulsation models to the observed
frequencies of the most thoroughly characterized helium-atmosphere
pulsator, GD~358. The free parameters in this initial study included the
stellar mass, the effective temperature, the surface helium layer mass,
the core composition, and the internal chemical profile.

For many years, astronomers have promised that the study of pulsating
white dwarfs would ultimately lead to useful information about the physics
of matter under extreme conditions of temperature and pressure. The
optimization approach developed in this dissertation has allowed us to
finally make good on that promise by exploiting the sensitivity of our
models to the core composition. We empirically determine that the central
oxygen abundance in GD~358 is $84 \pm 3$ percent. We use this value to
place a preliminary constraint on the $^{12}$C$(\alpha ,\gamma )^{16}$O
nuclear reaction cross-section of $S_{300}=295\pm15$ keV barns.

We find a thick helium-layer solution for GD~358 that provides a better
match to the data than previous fits, and helps to resolve a problem with
the evolutionary connection between PG~1159 stars and DBVs. We show that
the pulsation modes of our best-fit model probe down to the inner few
percent of the stellar mass. We demonstrate the feasibility of
reconstructing the internal chemical profiles of white dwarfs from
asteroseismological data, and we find an oxygen profile for GD~358 that is
qualitatively similar to recent theoretical calculations. This method
promises to serve as a powerful diagnostic that will eventually allow us to 
test theories of convective overshooting and stellar crystallization.

\vfill\cleardoublepage

\tableofcontents     
\vfill\cleardoublepage

\addcontentsline{toc}{chapter}{List of Tables}
\listoftables        
\vfill\cleardoublepage

\addcontentsline{toc}{chapter}{List of Figures}
\listoffigures       
\vfill\cleardoublepage

\pagestyle{headings}
\pagenumbering{arabic}

\chapter{Context}

\begin{quote}
``In learning any subject of a technical nature where mathematics plays a
role...it is easy to confuse the proof itself with the relationship it
establishes. Clearly, the important thing to learn and to remember is the
relationship, not the proof.''

\hskip 2.5in{\it ---Richard \citeauthor{fey63}}
\end{quote}

\section{Introduction}

There isn't much point in writing a dissertation if only a few people in
the world understand it, much less care enough about the subject to read
every word. It takes a long time to be educated as an astronomer, and by
the time it's over most students have internalized the basic concepts.
It's easy to forget the mental hurdles that challenged us along the way.

I began formal study in astronomy about ten years ago, and at every step
along the way my education has been subsidized by taxpayers. It seems only
fair that I should try to give something in return. I've decided to use
the first chapter of my dissertation to place my research project into a
larger social context. I will do my best to ensure that this chapter is
comprehensible to the people who so graciously and unknowingly helped me
along the way. It's the least I can do, and maybe it will convince some of
them that their investment was worthwhile.

\section{What Good is Astronomy?}

How can astronomers justify the support they receive from taxpayers? What
benefit does society derive from astronomical research? What is the rate
of return on their investment? These questions are difficult, but not
impossible, to answer. The economic benefits of basic scientific research
are often realized over the long-term, and it's hard to anticipate what
spin-offs may develop as a result of any specific research program.

A purist might refuse even to respond to these questions. What
justification does basic research need other than the pursuit of
knowledge? What higher achievement can civilization hope to accomplish
than the luxury of seeking answers to some of the oldest questions: where
did we come from, and what is our place in the universe?

The proper response is probably somewhere in between.  Over the years,
advances in scientific knowledge made possible through basic research have
had a definite impact on the average citizen, but the magnitude of this
impact is difficult to predict at the time the research is proposed. As a
result, much of the basic research funded by the public often sounds
ridiculous to many taxpayers. Gradually, this has led to a growing
reluctance by the public to fund basic research, and the annual budgets of
government funding agencies have stagnated as a consequence. Public
education is an essential component of any strategy to treat this problem
effectively.

I contend that the money taxpayers contribute to scientific research in
some sense obligates the researchers to make their work accessible to the
public. Some combination of teaching and public outreach by researchers
should provide an adequate return on the investment. If this doesn't seem
reasonable, put it in perspective by looking at exactly how much it costs
U.S. taxpayers to fund astronomical research: 
\begin{itemize}

\item In the year 2000, the total federal budget amounted to \$1.88
      trillion\footnotemark[1]

\footnotetext[1]{http://w3.access.gpo.gov/usbudget/fy2000/table2\_1.gif}

\item Revenue from personal income taxes amounted to \$900
      billion\footnotemark[1]

\item The ``non-defense discretionary'' portion of the budget totaled
      \$330 billion\footnotemark[2]

\footnotetext[2]{http://w3.access.gpo.gov/usbudget/fy2000/table2\_2.gif}

\item The National Science Foundation (NSF) budget was \$3.95
      billion\footnotemark[3]

\footnotetext[3]{http://www.nsf.gov/bfa/bud/fy2000/overview.htm}

\item The NSF allocation to the Directorate for Mathematics \&
      Physical Sciences (MPS) amounted to \$754 million\footnotemark[4]

\footnotetext[4]{http://www.nsf.gov/bfa/bud/fy2000/DirFund/mps.htm}

\item The MPS allocation to Astronomical Sciences amounted to \$122
      million\footnotemark[4]

\end{itemize}
So, even if you assume that the revenue for ``non-defense discretionary''
comes entirely from personal income taxes, funding astronomy is cheap. Out
of every \$1000 in revenue from personal income taxes, \$365 goes into the
non-defense discretionary fund. About \$4.35 ends up in the hands of the
National Science Foundation. Of this, 83 cents goes to fund all of
Mathematics \& Physical Sciences. In the end, for every \$1000 in taxes
only 13 cents ends up funding Astronomical Sciences.

\section{The Nature of Knowledge}

Science is chiefly concerned with accumulating {\it knowledge}. What does
this mean? The Greek philosopher Plato defined knowledge as ``justified
true belief''. Belief by itself is what we commonly call faith. There's
nothing wrong with faith, but it doesn't constitute knowledge under
Plato's definition. A belief that is justified but false is simply a
misconception. Based on incomplete information I may be justified in
believing that the Earth is flat, but I cannot {\it know} this to be so
because it turns out not to be true. Likewise, I may believe something
that turns out to be true even though I had no justification for believing
it. For example, I cannot {\it know} that a fair coin toss will turn up
heads even if it does in fact turn up heads, because I can have no
defensible justification for this belief.

In science, our justification for believing something is usually based on
observations of the world around us. The observations can occur either
before or after we have formulated a belief, corresponding to two broad
methods of reasoning. In {\it deductive} reasoning, we begin by
formulating a theory and deriving specific hypothetical consequences that
can be tested. We gather observations to test the hypotheses and help to
either confirm or refute the theory. In most cases the match between
observations and theory is imperfect, and we refine the theory to try to
account for the differences. Einstein's theory of relativity is a good
example of this type of reasoning. Based on some reasonable fundamental
assumptions, Einstein developed a theory of the geometry of the universe.
He predicted some observational consequences of this theory and people
tested these predictions experimentally.

For {\it inductive} reasoning, we begin by looking for patterns in
existing observations. We come up with some tentative hypotheses to
explain the patterns, and ultimately develop a general theory to explain
the observed phenomena. Kepler's laws of planetary motion are good
examples of inductive reasoning. Based on the precise observations of the
positions of planets in the night sky made by Tycho Brahe, Kepler noticed
some regular patterns. He developed several empirical laws that helped us
to understand the complex motions of the planets, which ultimately
inspired Newton to develop a general theory of gravity.

Armed with these methods of developing and justifying our beliefs, we
slowly converge on the truth. However, it's important to realize that we
may never actually arrive at our goal. We may only be able to find better
approximations to the truth. In astronomy we do not have the luxury of
designing the experiments or manipulating the individual components, so
knowledge in the strict sense is even more difficult to obtain.  
Fortunately, the universe contains such a vast and diverse array of
phenomena that we have plenty to keep us occupied.

\section{The Essence of my Dissertation Project}

When I originally conceived of my dissertation project three years ago,
the title of my proposal was {\it Genetic-Algorithm-based Optimization of
White Dwarf Pulsation Models using an Intel/Linux Metacomputer}. That's
quite a mouthful. It's actually much less intimidating than it sounds at
first. Let me explain what this project is really about, one piece at a
time.

\subsection{Genetic Algorithms}

Given the nature of knowledge, astronomers generally need to do two things
to learn anything useful about the universe. First, we need to gather
quantitative observations of something in the sky, usually with a
telescope and some sophisticated electronic detection equipment.  Second,
we need to interpret the observations by trying to match them with a
mathematical model, using a computer.  The computer models have many
different parameters---sort of like knobs and switches that can be
adjusted---and each represents some aspect of the physical laws that
govern the behavior of the model.

When we find a model that seems to match the observations fairly well, we
assume that the values of the parameters tell us something about the true
nature of the object we observed. The problem is: how do we know that some
other combination of parameters won't do just as well, or even better,
than the combination we found? Or what if the model is simply inadequate
to describe the true nature of the object?

The process of adjusting the parameters to find a ``best-fit'' model to
the observations is essentially an optimization problem. There are many
well established mathematical tools (algorithms) for doing this---each
with strengths and weaknesses. I am using a relatively new approach that
uses a process analogous to Charles Darwin's idea of evolution through
natural selection. This so-called {\it genetic algorithm} explores the
many possible combinations of parameters, and finds the best combination
based on objective criteria.

\subsection{White Dwarf Stars}

What is a white dwarf star? To astronomers, {\it dwarf} is a general term
for smaller stars. The color of a star is an indication of the temperature
at its surface. Very hot objects emit more blue-white light, while cooler
objects emit more red light. Our Sun is termed a {\it yellow dwarf} and
there are many stars cooler than the Sun called {\it red dwarfs}. So a
white dwarf is a relatively small star with a very hot surface.

In 1844, an astronomer named Friedrich Bessel noticed that Sirius, the
brightest star in the sky, appeared to wobble slightly as it moved through
space. He inferred that there must be something in orbit around it. Sure
enough, in 1862 the faint companion was observed visually by Alvan Clark
(a telescope maker) and was given the name ``Sirius B''. By the 1920's,
the companion had completed one full orbit of Sirius and its mass was
calculated, using Newton's laws, to be roughly the same as the Sun. When
astronomers measured its spectrum, they found that it emitted much more
blue light than red, implying that it was very hot on the surface even
though it didn't appear very bright in the sky. These observations implied
that it had to be a million times smaller than a regular star with the
same mass as the Sun---the first white dwarf!

The exact process of a star becoming a white dwarf depends on the mass of
the star, but all stars less massive than about 8 times the mass of the
Sun (99\% of all stars) will eventually become white dwarfs. Normal stars
fuse hydrogen into helium until the hydrogen deep in the center begins to
run out. For very massive stars this may take only 1 million years---but
for stars like the Sun the hydrogen lasts for 10,000 million years. When
enough helium collects in the middle of the star, it becomes a significant
source of extra heat. This messes up the internal balance of the star,
which then begins to bloat into a so-called red giant.

\begin{figure}
\hskip 0.1in
\epsfxsize 5.5in
\epsffile{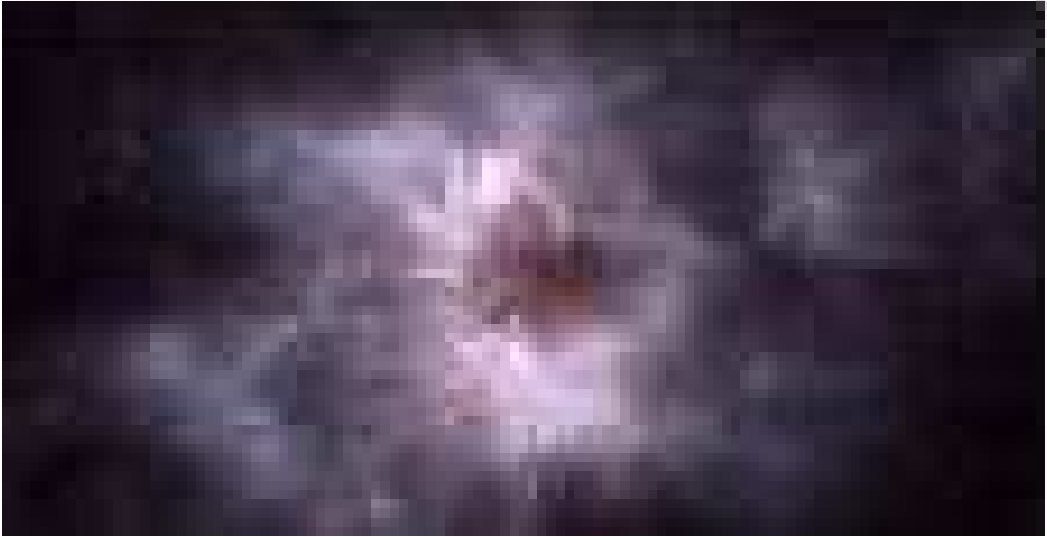}
\caption[Hubble Space Telescope image of a hot white dwarf]
{A hot white dwarf at the center of the planetary nebula NGC 2440. This
image was obtained with the Hubble Space Telescope by H. Bond (STScI) and
R. Ciardullo (PSU).\label{fig1.1}}
\vspace*{-6pt}
\end{figure}

If the star is massive enough, it may eventually get hot enough in the
center to fuse the helium into carbon and oxygen. The star then enjoys
another relatively stable period, though much shorter this time. The
carbon and oxygen, in their turn, collect in the middle. If the star isn't
massive enough to reach the temperature needed to fuse carbon and oxygen
into heavier elements, then these elements will simply continue to collect
in the center until the helium fuel runs out. In the end, you have a
carbon/oxygen white dwarf surrounded by the remains of the original star
(see Figure \ref{fig1.1}).

% FIGURE 1.1 %%%%%%%%%%%%%%%%%%%%%%%%%%%%%%%%%%%%%%%%%%%%%%%%%%%%%%%%%%%%%%
%%%%%%%%%%%%%%%%%%%%%%%%%%%%%%%%%%%%%%%%%%%%%%%%%%%%%%%%%%%%%%%%%%%%%%%%%%%

In normal stars like the Sun, the inward pull of gravity is balanced by
the outward push of the high-temperature material in the center, fusing
hydrogen into helium and releasing energy in the process. There is no
nuclear fusion in a white dwarf. Instead, the force that opposes gravity
is called ``electron degeneracy pressure''.

When electrons are squeezed very close together, the energy-states that
they would normally be able to occupy become indistinguishable from the
energy-states of neighboring electrons. The rules of quantum mechanics
tell us that no two electrons can occupy exactly the same energy-state,
and as the average distance between electrons gets smaller the average
momentum must get larger. So, the electrons are forced into higher
energy-states (pushed to higher speeds) just because of the density of the
matter.

% FIGURE 1.2 %%%%%%%%%%%%%%%%%%%%%%%%%%%%%%%%%%%%%%%%%%%%%%%%%%%%%%%%%%%%%%
\begin{figure}
\epsfxsize 5.5in
\epsffile{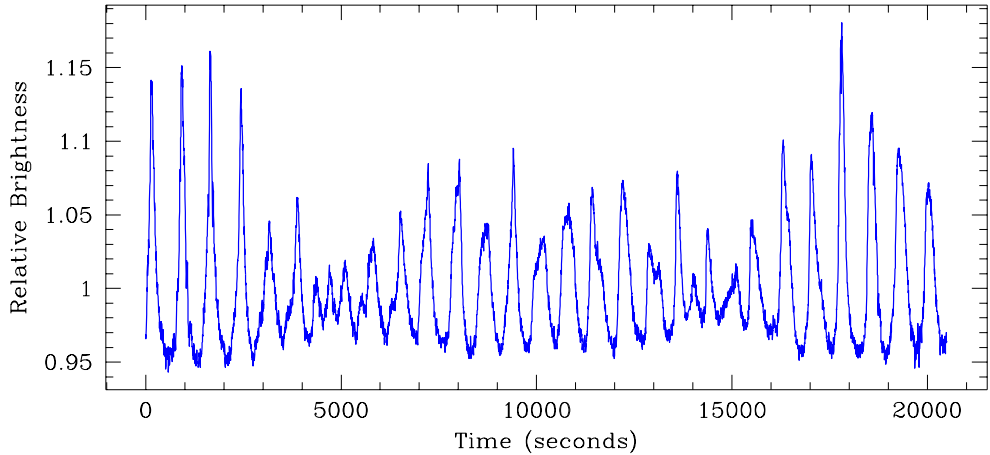}
\caption[Sample light curve of the pulsating white dwarf GD~358]
{Non-radial pulsations in some white dwarf stars cause periodic changes in
their brightness over time. This sample light curve of GD~358 shows many
pulsation periods excited simultaneously, causing beating and a total
variation of about 10 percent on timescales shorter than an hour.
\label{fig1.2}}
\end{figure}
%%%%%%%%%%%%%%%%%%%%%%%%%%%%%%%%%%%%%%%%%%%%%%%%%%%%%%%%%%%%%%%%%%%%%%%%%%%

This quantum pressure can oppose gravity as long as the density doesn't
get too high. If a white dwarf has more than about 1.4 times the mass of
the Sun squeezing the material, there will be too few energy-states
available to the electrons (since they cannot travel faster than the speed
of light) and the star will collapse---causing a supernova explosion.

\vspace*{-6pt}
\subsubsection{Pulsating White Dwarfs}

Some white dwarfs show very regular variations in the amount of light
reaching our telescopes (see Figure \ref{fig1.2}). The pattern of this
variation suggests that these white dwarfs are pulsating---as if there are
continuous star-quakes going on. By studying the patterns of light
variation, astronomers can learn about the interior structure of white
dwarfs---in much the same way as seismologists can learn about the inside
of the Earth by studying earthquakes. For this reason, the study of these
pulsating white dwarfs is called asteroseismology.

Since 1988, very useful observations of pulsating white dwarfs have been
obtained with the Whole Earth Telescope---a collaboration of astronomers
around the globe who cooperate to monitor these stars for weeks at a time.
I have helped to make some of these observations, but I have also worked
on interpreting them using our computer models. I have approached the
models in two ways:

\begin{itemize}

\item I assume the models are accurate representations of the real white
      dwarf stars, and I try to find the combination of model parameters 
      that do the best job of matching the observations.

\item I assume the models are incomplete representations of the real white
      dwarf stars, and I try to find changes to the internal structure of 
      the models that yield an improved match to the observations.

\end{itemize}

\subsection{Linux Metacomputer}

The dictionary definition of the prefix meta- is: ``Beyond; More
comprehensive; More highly developed.'' So a meta-computer goes beyond the
boundaries of a traditional computer as we are accustomed to thinking of
it. Essentially, a metacomputer is a collection of many individual
computers, connected by a network (the Internet for example), which can
cooperate on solving a problem. In general, this allows the problem to be
solved much more quickly than would be possible using a single computer.

Supercomputers are much faster than a single desktop computer too, but
they usually cost millions of dollars, and everyone has to compete for
time to work on their problem. Recently, personal computers have become
very fast and relatively inexpensive. At the same time, the idea of free
software (like the Linux operating system) has started to catch on. These
developments have made it feasible to build a specialized metacomputer
with as much computing power as a 5-year-old supercomputer, but for only
about 1\% of the cost!

\begin{figure}[b]
\vspace*{-12pt}
\hskip 0.1in
\epsfxsize 5.5in
\epsffile{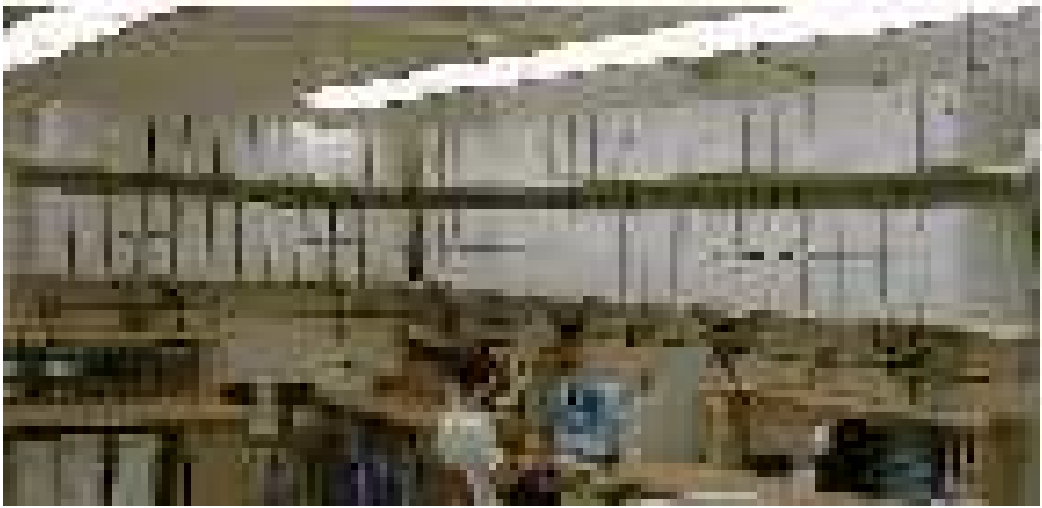}
\caption[The parallel computer built and used for this thesis]
{This metacomputer is a collection of 64 minimal PCs connected by a
network, and can calculate our white dwarf models in parallel.  
\label{fig1.3}}
\end{figure}

The problem that I am working on has required that I run literally
millions of computer models of pulsating white dwarf stars over the
several-year duration of my research project. To make these calculations
practical, I configured a metacomputer using 64 minimal PC systems running
under a customized version of the Linux operating system (see Figure
\ref{fig1.3}).

% FIGURE 1.3 %%%%%%%%%%%%%%%%%%%%%%%%%%%%%%%%%%%%%%%%%%%%%%%%%%%%%%%%%%%%%%
%%%%%%%%%%%%%%%%%%%%%%%%%%%%%%%%%%%%%%%%%%%%%%%%%%%%%%%%%%%%%%%%%%%%%%%%%%%

Thanks to another piece of free software called PVM (for Parallel Virtual
Machine), I can use one fully-equipped personal computer to control the
entire system. This central computer is responsible for distributing work
to each of the 64 processors, and collecting the results. There is a small
amount of work required just to keep track of everything, so the
metacomputer actually runs about 60 (rather than 64) times as fast as a
single system. Not bad!

\vspace*{-0.1in}
\subsection{The Big Picture}

So I'm using a relatively new optimization method to find the best set of
parameters to match the observations with our computer models of pulsating
white dwarf stars; and there are so many models to run that I need a lot
of computing power, so I linked a bunch of PC systems together to do the
job. But what do I hope to learn?

Well, the source of energy for regular stars like the Sun is nuclear
fusion.  This is the kind of nuclear energy that doesn't produce any
dangerous radioactive waste. Astronomers have a good idea of how fusion
energy works to power stars, but the process requires extremely high
temperatures and pressures, which must be sustained for a long time; these
conditions are difficult to reproduce (in a controlled way) in
laboratories on Earth. Physicists have been working on it for several
decades, but sustained nuclear fusion has still never been achieved. This
leads us to believe that we may not understand all of the physics that we
need to make fusion work. If scientists could learn to achieve
controlled nuclear fusion, it would provide an essentially inexhaustible
source of clean, sustainable energy.

To help ensure that we properly understand how stars work, it is useful to
look at the ``ashes'' of the nuclear fusion. Those ashes are locked in the
white dwarf stars, and asteroseismology allows us to peer down inside and
probe around. But our understanding can only be as good as our models, so
it is important both to make sure that we find the absolute ``best''
match, and to figure out what limitations are imposed simply by using the
models we use. It's only one piece of the puzzle, but it's a place to
start.

\vspace*{-0.1in}
\section{Organization of this Dissertation}
\vspace*{-0.05in}

Most of the work presented in this dissertation has already been
published. Each chapter should be able to stand by itself, and together
they tell the story of how I've spent the last three years of my
professional life.

Chapter \ref{meta} describes the development of the Linux metacomputer and
provides a detailed account of its inner workings. The text is derived
from articles published in the Linux Journal, Baltic Astronomy, and a
Metacomputer mini-{\sc howto} posted on the world-wide web.

Chapter \ref{ga} includes a more detailed background on genetic
algorithms. I outline the steps I took to create a parallel version of a
general-purpose genetic algorithm in the public domain, and its
implementation on the Linux metacomputer.

Chapter \ref{fwd} is derived primarily from a paper published in the 
Astrophysical Journal in December 2000. It describes the first application 
of the genetic algorithm approach to model pulsations in the white dwarf 
GD~358, and an extension of the method to determine its internal 
composition and structure.

Chapter \ref{rev} comes from a paper published in the Astrophysical
Journal in August 2001. It describes a method of ``reverse engineering''
the internal structure of a pulsating white dwarf by using the genetic
algorithm and the models in a slightly different way.

Chapter \ref{conc} sums up the major conclusions of this work and outlines
future directions. The appendices contain an archive of my observations
for the Whole Earth Telescope, some interactive simulations of pulsating
white dwarfs, and an archive of the computer codes used for this
dissertation.

\vfill\cleardoublepage\thispagestyle{empty}

\chapter{Linux Metacomputer \label{meta}}

\begin{quote}
``There's certainly a strong case for people disliking
Microsoft because their operating systems... suck.''

\hskip 2.5in{\it ---Linus \citeauthor{tor99}}
\end{quote}

\section{Introduction}

The adjustable parameters in our computer models of white dwarfs presently
include the total mass, the temperature, hydrogen and helium layer masses,
core composition, convective efficiency, and internal chemical profiles.  
Finding a proper set of these to provide a close fit to the observed data
is difficult. The traditional procedure is a guess-and-check process
guided by intuition and experience, and is far more subjective than we
would like.  Objective procedures for determining the best-fit model are
essential if asteroseismology is to become a widely-accepted and reliable
astronomical technique. We must be able to demonstrate that for a given
model, within the range of different values the model parameters can 
assume, we have found the only solution, or the best one if more than one 
is possible. To address this problem, we have applied a search-and-fit 
technique employing a genetic algorithm (GA), which can explore the myriad 
parameter combinations possible and select for us the best one, or ones
\citep[cf.][]{gol89,cha95,met99}.

\section{Motivation}

Although genetic algorithms are often more efficient than other 
global techniques, they are still quite demanding computationally. On a
reasonably fast computer, it takes about a minute to calculate the
pulsation periods of a single white dwarf model. However, finding the
best-fit with the GA method requires the evaluation of hundreds of
thousands of such models. On a single computer, it would take more than
two months to find an answer. To develop this method on a reasonable
timescale, we realized that we would need our own parallel computer.

It was January 1998, and the idea of parallel computing using inexpensive
personal computer (PC) hardware and the free Linux operating system
started getting a lot of attention. The basic idea was to connect a bunch
of PCs together on a network, and then to split up the computing workload
and use the machines collectively to solve the problem more quickly. Such
a machine is known to computer scientists as a {\it metacomputer}. This
differs from a {\it supercomputer}, which is much more expensive since all
of the computing power is integrated into a single unified piece of
hardware.

There are several advantages to using a metacomputer rather than a more
traditional supercomputer. The primary advantage is price: a metacomputer
that is just as fast as a 5-year-old supercomputer can be built for only
about 1 percent of the cost---about \$10,000 rather than \$1 million!
Another major advantage is access: the owner and administrator of a
parallel computer doesn't need to compete with other researchers for time
or resources, and the hardware and software configuration can be optimized
for a specific problem. Finally if something breaks, replacement parts are
standard off-the-shelf components that are widely available, and while
they are on order the computer is still functional at a slightly reduced
capacity.

\section{Hardware}

\footnotetext[5]{http://www.beowulf.org/}

The first Linux metacomputer, known as the Beowulf cluster\footnotemark[5]
\citep{bec95}, has now become the prototype for many general-purpose Linux 
clusters. Our machine is similar to Beowulf in the sense that it consists 
of many independent PCs, or {\it nodes}; but our goal was to design a 
special-purpose computational tool with the best performance possible per 
dollar, so our machine differs from Beowulf in several important ways.

\begin{figure}[b]
\hskip 0.1in
\epsfxsize 5.5in
\epsffile{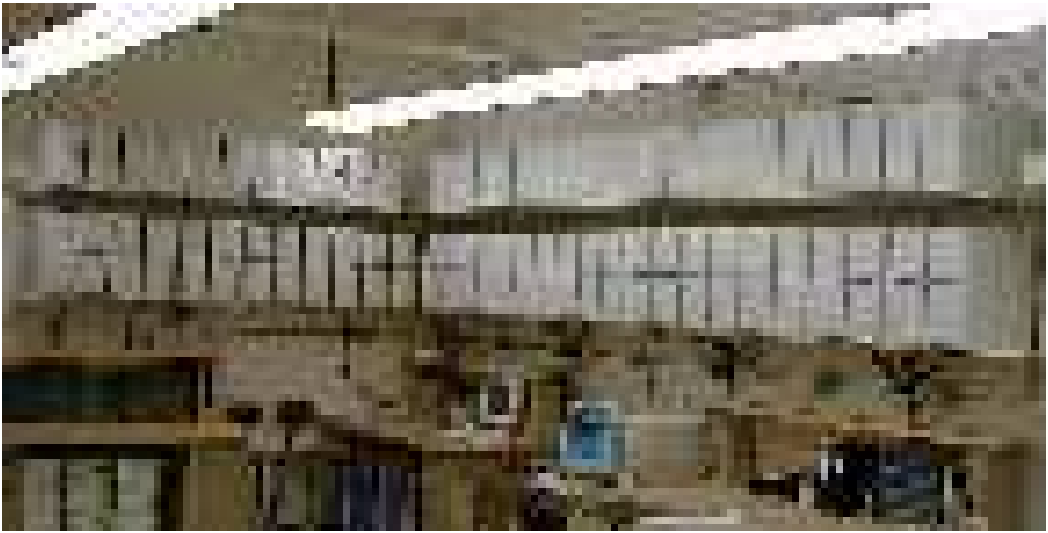}
\caption[The master computer surrounded by 64 slave nodes]
{The 64 minimal nodes of the metacomputer on shelves surrounding the
master computer.
\label{fig2.1}}
\end{figure}

We wanted to use each node of the metacomputer to run identical tasks
(white dwarf pulsation models) with small, independent sets of data (the
parameters for each model). The results of the calculations performed by
the nodes consisted of just a few numbers (the root-mean-square
differences between the observed and calculated pulsation periods) which
only needed to be communicated to the master process (the genetic
algorithm), never to another node. Essentially, network bandwidth was not
an issue because the computation to communication ratio of our application
was extremely high, and hard disks were not needed on the nodes because
our problem did not require any significant amount of data storage. We
settled on a design including one master computer and 64 minimal nodes
connected by a simple coaxial network (see Figure \ref{fig2.1}).

% FIGURE 2.1 %%%%%%%%%%%%%%%%%%%%%%%%%%%%%%%%%%%%%%%%%%%%%%%%%%%%%%%%%%%%%%
%%%%%%%%%%%%%%%%%%%%%%%%%%%%%%%%%%%%%%%%%%%%%%%%%%%%%%%%%%%%%%%%%%%%%%%%%%%

We developed the metacomputer in four phases. To demonstrate that we could
make the system work, we started with the master computer and only two
nodes. When the first phase was operational, we expanded it to a dozen
nodes to demonstrate that the performance would scale. In the third phase,
we occupied the entire bottom shelf with a total of 32 nodes. Months
later, we were given the opportunity to expand the system by an additional
32 nodes with processors donated by AMD and we filled the top shelf,
yielding a total of 64 nodes.

\subsection{Master Computer}

\begin{figure}[b]
\hskip 0.1in
\epsfxsize 5.5in
\epsffile{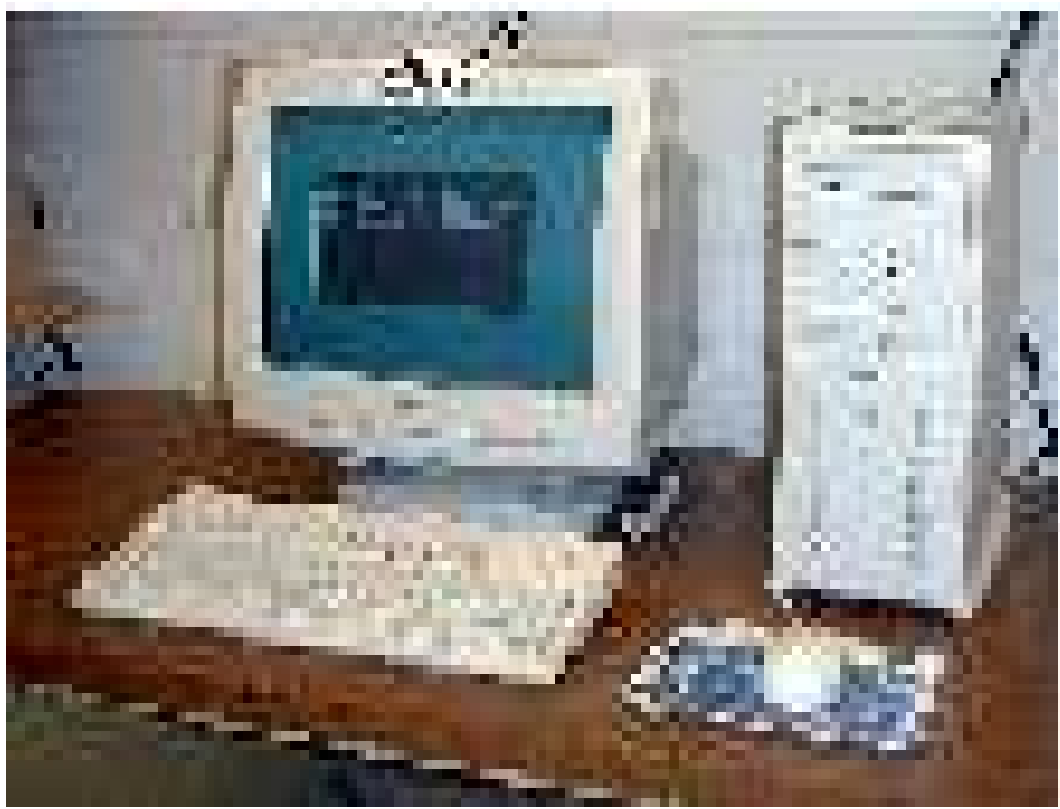}
\caption[Darwin, the master computer.]
{Darwin, the master computer controlling all 64 nodes.\label{fig2.2}}
\end{figure}

Our master computer, which we call Darwin, is a Pentium-II 333 MHz system
with 128 MB RAM and two 8.4 GB hard disks (see Figure \ref{fig2.2}). It
has three NE-2000 compatible network cards, each of which drives 1/3 of
the nodes on a subnet. No more than 30 devices (e.g. ethernet cards in the
nodes) can be included on a single subnet without using a repeater to
boost the signal. Additional ethernet cards for the master computer were
significantly less expensive than a repeater.

% FIGURE 2.2 %%%%%%%%%%%%%%%%%%%%%%%%%%%%%%%%%%%%%%%%%%%%%%%%%%%%%%%%%%%%%
%%%%%%%%%%%%%%%%%%%%%%%%%%%%%%%%%%%%%%%%%%%%%%%%%%%%%%%%%%%%%%%%%%%%%%%%%%%

\subsection{Slave Nodes}

\begin{figure}[b]
\hskip 0.1in
\epsfxsize 5.5in
\epsffile{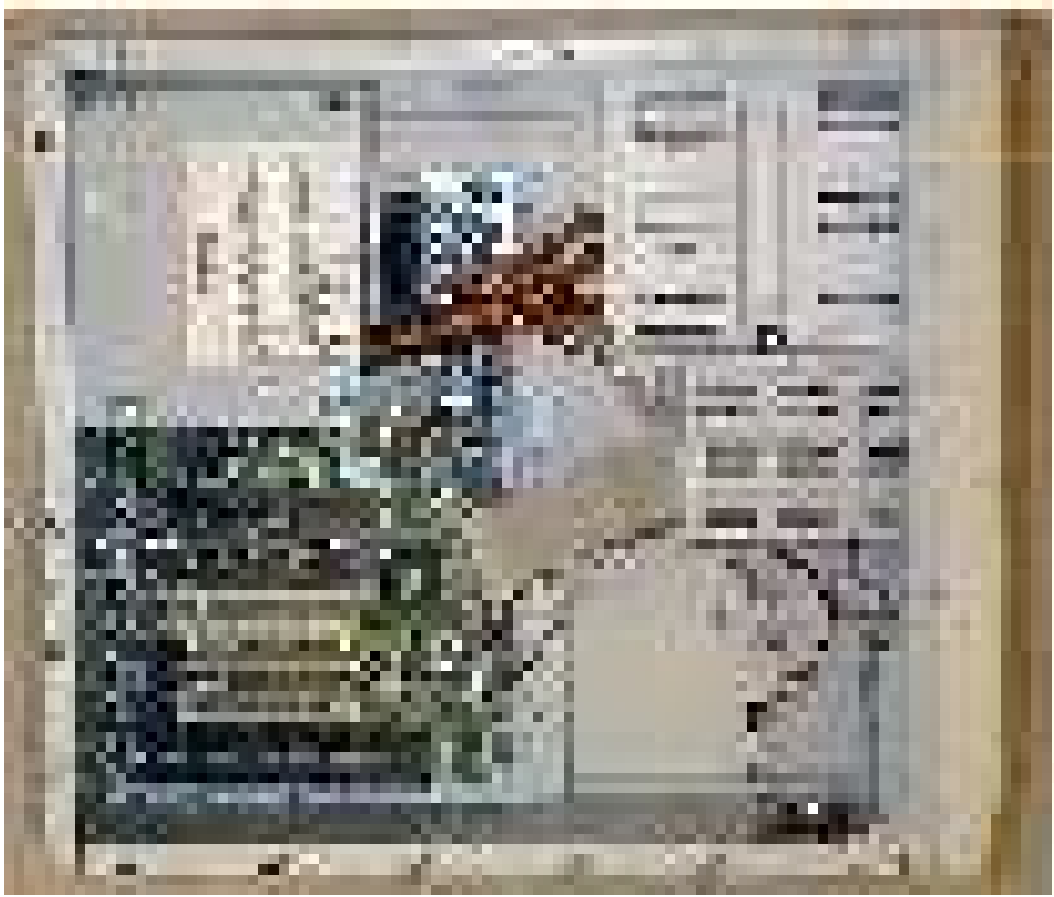}
\caption[A view inside one of the metacomputer nodes]
{A view inside one of the metacomputer nodes.\label{fig2.3}}
\end{figure}

We assembled the nodes from components obtained at a local discount
computer outlet. Each node includes only an ATX tower case and power
supply with a motherboard, a processor and fan, a single 32 MB RAM chip,
and an NE-2000 compatible network card (see Figure \ref{fig2.3}). Half of
the nodes contain Pentium-II 300 MHz processors, while the other half are
AMD K6-II 450 MHz chips. We added inexpensive Am27C256 32 kb EPROMs
(erasable programmable read-only memory) to the bootrom sockets of each
network card. The nodes are connected in series with 3-ft ethernet coaxial
cables, and the subnets have 50 $\Omega$ terminators on each end. The
total cost of the system was around \$25,000 but it could be built for
considerably less today, and less still tomorrow.

% FIGURE 2.3 %%%%%%%%%%%%%%%%%%%%%%%%%%%%%%%%%%%%%%%%%%%%%%%%%%%%%%%%%%%%%
%%%%%%%%%%%%%%%%%%%%%%%%%%%%%%%%%%%%%%%%%%%%%%%%%%%%%%%%%%%%%%%%%%%%%%%%%%%

\section{Software}

Configuring the software was not much more complicated than setting up a
diskless Linux system. The main difference was that we wanted to minimize
network activity by giving each node an identical, independent filesystem
rather than mounting a shared network filesystem. Since the nodes had no
hard disks, we needed to create a self-contained filesystem that could be
downloaded and mounted in a modest fraction of the 32 MB RAM.

To make the system work, we relied heavily on the open-source Linux
operating system and other software packages that were available for free
on the Internet. A piece of software called YARD allowed us to create the
minimal Linux filesystems that we needed for each node to run
independently.  We used a package called NETBOOT to program the EPROM
chips; this allowed each node to automatically download and boot the
filesystem, which was mounted in part of the RAM. Finally, we used the PVM
software to exploit the available resources and take care of the
communication required to manage the parallel processing operations.

\subsection{Linux}

In 1991, a young student named Linus Torvalds at the University of
Helsinki in Finland created a free Unix-like operating system as a hobby.
He posted his work on the Internet and, together with a small group of
friends, he continued to develop it. In 1994, version 1.0 of ``Linux'' was
released.  Today, millions of people worldwide use Linux as an alternative
to the operating systems sold by Microsoft (Windows) and Sun Microsystems
(Solaris). Unlike these more common operating systems, the source code for
Linux is freely available to everyone.

The computer code used to create the Linux operating system is known as
the {\it kernel}. To ensure that the hardware components of our nodes
would be recognized by the operating system, we custom compiled the Linux
2.0.34 kernel. We included support for the NE-2000 ethernet card, and
specified that the filesystem was on the network and should be retrieved
using the {\tt bootp} protocol (see below) and mounted in RAM.

Getting the master computer to recognize its three ethernet cards required
extra options to be passed to the kernel at boot time. We specified the
addresses of these devices and passed them to the Linux kernel through
LOADLIN, a DOS-based program that boots up Linux.

Each card on the network needed to be assigned a unique IP (Internet
Protocol) address, which is a sequence of four numbers between 0 and 255
separated by periods. The IP addresses that are reserved for subnets
(which do not operate on the Internet) are:
\begin{center}
\begin{tabular}{lr}
        10.0.0.0                                 & (Class A network) \\
        172.16.0.0  $\rightarrow$ 172.31.0.0     & (Class B network) \\
        192.168.0.0 $\rightarrow$ 192.168.255.0  & (Class C network) \\
\end{tabular}
\end{center}
Since we were dealing with a relatively small number of machines, we used
the first three numbers to specify the domain (pvm.net), and the last
number to specify the hostname (e.g. node001). Our first ethernet card
({\tt eth1})  was assigned control of the 192.168.1.0 subnet while
192.168.2.0 and 192.168.3.0 were handled by {\tt eth2} and {\tt eth3}
respectively.

We used the Bootstrap Protocol ({\tt bootp}) and the Trivial File Transfer
Protocol ({\tt tftp}) to allow the nodes to retrieve and boot their
kernel, and to download a compressed version of their root filesystem. We
relied heavily on Robert Nemkin's Diskless HOWTO\footnotemark[6] to make 
it work.

\footnotetext[6]{http://www.linuxdoc.org/HOWTO/Diskless-HOWTO.html}

The main configuration file for {\tt bootp} is {\tt /etc/bootptab}, which
contains a list of the hostnames and IP addresses that correspond to each
ethernet card on the subnet. Each card is identified by a unique hardware
address---a series of 12 hexadecimal numbers (0-9,a-f) assigned by the
manufacturer. In addition to various network configuration parameters,
this file also describes the location of the bootimage to retrieve with
{\tt tftp}. Since each node is running an identical copy of the bootimage,
setting up {\tt tftp} was considerably easier than it would have been in
general. We simply created a {\tt /tftpboot} directory on the server and
placed a copy of the bootimage there.

\subsection{YARD}

To create the self-contained root filesystem, we used Tom Fawcett's YARD
(Yet Another Rescue Disk) package\footnotemark[7]. This piece of software
was designed to make rescue disks---self-contained minimal filesystems
that can fit on a single floppy disk and be used in emergencies to boot
and fix problems on a Linux system. Since the white dwarf pulsation code
does not require a great deal of system memory to run, we were free to use
half of the 32 MB RAM for our filesystem, which allowed us to include much
more than would fit on a floppy disk.

\footnotetext[7]{http://www.croftj.net/$\sim$fawcett/yard/}

There are two files that control the properties and content of the YARD
filesystem: {\tt Config.pl} and {\tt Bootdisk\_Contents}. The {\tt
Config.pl} file controls the size of the filesystem, the location of the
kernel image, and other logistical matters. The {\tt Bootdisk\_Contents}
file contains a list of the daemons, devices, directories, files,
executable programs, libraries, and utilities that we explicitly wanted to
include in the filesystem. The scripts that come with YARD automatically
determine the external dependences of anything included, and add those to
the filesystem before compressing the whole thing.

\subsection{NETBOOT}

We used Gero Kuhlmann's NETBOOT package\footnotemark[8]\ to create the
bootimage that each node downloads from the master computer. The bootimage
is really just a concatenated copy of the Linux kernel ({\tt zImage.node})  
and the compressed root filesystem ({\tt root.gz}). The NETBOOT software
also includes a utility for creating a ROM image that is used to program
the EPROMs in the ethernet card for each node. Although our ROM image was
only 16 kb, we used Am27C256 (32 kb) EPROMs because they were actually
cheaper than the smaller chips.

\footnotetext[8]{http://www.han.de/$\sim$gero/netboot/}

\subsection{PVM}

\footnotetext[9]{http://www.epm.ornl.gov/pvm/} 

The Parallel Virtual Machine (PVM) software\footnotemark[9]\ allows a
collection of computers connected by a network to cooperate on a problem
as if they were a single multi-processor parallel machine. It was
developed in the 1990's at Oak Ridge National Laboratory \citep{gei94}.
The software consists of a daemon running on each host in the virtual
machine, and a library of routines that need to be incorporated into a
computer program so that it can utilize all of the available computing
power.

\section{How it works}

With the master computer up and running, we turn on one node at a time (to
prevent the server from being overwhelmed by many simultaneous {\tt bootp}
requests). By default, the node tries to boot from the network first. It
finds the bootrom on the ethernet card, and executes the ROM program. This
program initializes the ethernet card and broadcasts a {\tt bootp} request
over the network.

When the server receives the request, it identifies the unique hardware
address, assigns the corresponding IP address from the {\tt /etc/bootptab}
file, and allows the requesting node to download the bootimage. The node
loads the Linux kernel image into memory, creates a 16 MB initial ramdisk,
mounts the root filesystem, and starts all essential services and daemons.

Once all of the nodes are up, we start the PVM daemons on each node from
the master computer. Any computer program that incorporates the PVM
library routines and has been included in the root filesystem can then be
run in parallel.

\section{Benchmarks}

Measuring the absolute performance of the metacomputer is difficult
because the result strongly depends on the fraction of Floating-point
Division operations (FDIVs) used in the benchmark code. Table \ref{tab2.1}
lists four different measures of the absolute speed in Millions of
FLoating-point Operations Per Second (MFLOPS).

% TABLE 2.1 %%%%%%%%%%%%%%%%%%%%%%%%%%%%%%%%%%%%%%%%%%%%%%%%%%%%%%
\begin{table}
\begin{center}
\caption{\label{tab2.1}The Absolute Speed of the Metacomputer}
\vskip 5pt
\begin{tabular}{cccc}\hline\hline
Benchmark & P-II 300 MHz & K6-II 450 MHz & Total Speed \\
\hline
MFLOPS(1) & 80.6         & ~65.1         & 4662.4      \\
MFLOPS(2) & 47.9         & ~67.7         & 3699.2      \\
MFLOPS(3) & 56.8         & 106.9         & 7056.0      \\
MFLOPS(4) & 65.5         & 158.9         & 7180.8      \\
\hline\hline
\end{tabular}
\vskip -8pt
\end{center}
\end{table}
%%%%%%%%%%%%%%%%%%%%%%%%%%%%%%%%%%%%%%%%%%%%%%%%%%%%%%%%%%%%%%%%%%

The code for MFLOPS(1) is essentially scalar, which means that it cannot
exploit any advantages that are intrinsic to processor instruction sets;
the percentage of FDIVs (9.6\%) is considered somewhat high. The code for
MFLOPS(2) is fully vectorizable, which means that it can exploit
advantages intrinsic to each processor, but the percentage of FDIVs
(9.2\%)  is still on the high side. The code for MFLOPS(3) is also fully
vectorizable and the percentage of FDIVs (3.4\%) is considered moderate.
The code for MFLOPS(4) is fully vectorizable, but the percentage of FDIVs
is zero. We feel that MFLOPS(3) provides the best measure of the expected
performance for the white dwarf code because of the moderate percentage
of FDIVs. Adopting this value, we have achieved a price to performance
ratio near \$3.50/MFLOPS.

The relative speed of the metacomputer is easy to measure. We simply
compare the amount of time required to compute a specified number of white
dwarf models using all 64 nodes to the amount of time required to
calculate the same number of models using only one of the nodes. We find
that the metacomputer is about 60 times faster than a single node by
itself.

\section{Stumbling Blocks}

After more than 3 months without incident, one of the nodes abruptly died.
One of the graduate students working in our lab reported, ``One of your
babies is crying!'' As it turned out, the power supply had gone bad,
frying the motherboard and the CPU fan. The processor overheated, shut
itself off, and triggered an alarm. We now keep a few spare CPU fans and
power supplies on hand. This is the only real problem we have had with the
system, and it was easily fixed.

Since the first incident, this scenario has repeated itself five times
over a three year period. This implies that such events can be expected at
the rate of 2 per year for this many nodes. In addition to the more
serious failures, there have been ten other power supply failures which
did not result in peripheral hardware damage. The rate for these failures
is 3-4 per year.

\vfill\cleardoublepage\thispagestyle{empty}

\chapter{Parallel Genetic Algorithm \label{ga}}

\begin{quote}
``Evolution is cleverer than you are.''

\hskip 2.5in{\it ---Francis \citeauthor{cri84}}
\end{quote}

\section{Background}

The problem of extracting useful information from a set of observational
data often reduces to finding the set of parameters for some theoretical
model which results in the closest match to the observations. If the
physical basis of the model is both accurate and complete, then the values
of the parameters for the best-fit model can yield important insights into
the nature of the object under investigation.

When searching for the best-fit set of parameters, the most fundamental
consideration is: where to begin? Models of all but the simplest physical
systems are typically non-linear, so finding the least-squares fit to the
data requires an initial guess for each parameter. Generally, some
iterative procedure is used to improve upon this first guess in order to
find the model with the absolute minimum residuals in the
multi-dimensional parameter-space.

There are at least two potential problems with this standard approach to
model fitting. The initial set of parameters is typically determined by
drawing upon the past experience of the person who is fitting the model.  
This {\it subjective} method is particularly disturbing when combined with
a {\it local} approach to iterative improvement.  Many optimization
schemes, such as differential corrections \citep{pl72} or the simplex
method \citep{kl87}, yield final results which depend to some extent on
the initial guesses. The consequences of this sort of behavior are not
serious if the parameter-space is well behaved---that is, if it contains a
single, well defined minimum. If the parameter-space contains many local
minima, then it can be more difficult for the traditional approach to find
the global minimum.

\section{Genetic Algorithms}

An optimization scheme based on a genetic algorithm (GA) can avoid the
problems inherent in more traditional approaches. Restrictions on the
range of the parameter-space are imposed only by observations and by the
physics of the model. Although the parameter-space so-defined is often
quite large, the GA provides a relatively efficient means of searching
globally for the best-fit model. While it is difficult for GAs to find
precise values for the set of best-fit parameters, they are well suited to
search for the {\it region} of parameter-space that contains the global
minimum. In this sense, the GA is an objective means of obtaining a good
first guess for a more traditional method which can narrow in on the
precise values and uncertainties of the best-fit.

The underlying ideas for genetic algorithms were inspired by Charles
Darwin's \citeyearpar{dar59} notion of biological evolution through
natural selection. The basic idea is to solve an optimization problem by
{\it evolving} the best solution from an initial set of completely random
guesses. The theoretical model provides the framework within which the
evolution takes place, and the individual parameters controlling it serve
as the genetic building blocks. Observations provide the selection
pressure. A comprehensive description of how to incorporate these ideas in
a computational setting was written by \cite{gol89}.

Initially, the parameter-space is filled uniformly with trials consisting
of randomly chosen values for each parameter, within a range based on the
physics that the parameter is supposed to describe. The model is evaluated
for each trial, and the result is compared to the observed data and
assigned a {\it fitness} based on the relative quality of the match. A new
generation of trials is then created by selecting from this population at
random, weighted by the fitness.

To apply genetic operations to the new generation of trials, their
characteristics must be encoded in some manner. The most straightforward
way of encoding them is to convert the numerical values of the parameters
into a long string of numbers. This string is analogous to a chromosome,
and each number represents a gene. For example, a two parameter trial with
numerical values $x_1=1.234$ and $y_1=5.678$ would be encoded into a
single string of numbers `12345678'.

Next, the encoded trials are paired up and modified in order to explore
new regions of parameter-space. Without this step, the final solution
could ultimately be no better than the single best trial contained in the
initial population. The two basic operations are {\it crossover} which
emulates sexual reproduction, and {\it mutation} which emulates
happenstance and cosmic rays.

As an example, suppose that the encoded trial above is paired up with
another trial having $x_2=2.468$ and $y_2=3.579$, which encodes to the
string `24683579'. The crossover procedure chooses a random position
between two numbers along the string, and swaps the two strings from that
position to the end. So if the third position is chosen, the strings
become
$$
\begin{array}{c}
123|45678 \rightarrow 123|83579 \\
246|83579 \rightarrow 246|45678
\end{array}
$$
Although there is a high probability of crossover, this operation is not
applied to all of the pairs. This helps to keep favorable characteristics
from being eliminated or corrupted too hastily. To this same end, the rate
of mutation is assigned a relatively low probability. This operation
allows for the spontaneous transformation of any particular position on
the string into a new randomly chosen value. So if the mutation operation
were applied to the sixth position of the second trial, the result might
be
$$
24645|6|78 \rightarrow 24645|0|78
$$

After these operations have been applied, the strings are decoded back
into sets of numerical values for the parameters. In this example, the new
first string `12383579' becomes $x_1=1.238$ and $y_1=3.579$ and the new
second string `24645078' becomes $x_2=2.464$ and $y_2=5.078$.  This new
generation replaces the old one, and the process begins again.  The
evolution continues until one region of parameter-space remains populated
while other regions become essentially empty. The robustness of the
solution can be established by running the GA several times with different
random initialization.

Genetic algorithms have been used a great deal for optimization problems
in other fields, but until recently they have not attracted much attention
in astronomy. The application of GAs to problems of astronomical interest
was promoted by \cite{cha95}, who demonstrated the technique by fitting
the rotation curves of galaxies, a multiply-periodic signal, and a
magneto-hydrodynamic wind model. Many other applications of GAs to
astronomical problems have appeared in the recent literature. \cite{hak95}
optimized the accretion stream map of an eclipsing polar. \cite{lan95}
developed an optimum set of image selection criteria for detecting
high-energy gamma rays. \cite{ken95} used radial velocity observations to
identify the oscillation modes of a $\delta$ Scuti star. \cite{laz97}
searched pulsar timing signals for the signatures of planetary companions.
\cite{cha98} performed a helioseismic inversion to constrain solar core
rotation. \cite{wah98} determined the orbital parameters of interacting
galaxies. \cite{met99} used a GA to fit the light curves of an eclipsing
binary star. The applicability of GAs to such a wide range of astronomical
problems is a testament to their versatility.

\section{Parallelizing PIKAIA}

There are only two ways to make a computer program run faster---either
make the code more efficient, or run it on a faster machine. We made a few
design improvements to the original white dwarf code, but they decreased
the runtime by only $\sim$10\%. We decided that we really needed access to
a faster machine. We looked into the supercomputing facilities available
through the university, but the idea of using a supercomputer didn't
appeal to us very much; the process seemed to involve a great deal of red
tape, and we weren't certain that we could justify time on a supercomputer
in any case. To be practical, the GA-based fitting technique required a
dedicated instrument to perform the calculations. We designed and
configured such an instrument---an isolated network of 64 minimal PCs
running Linux \citep{mn99,mn00}. To allow the white dwarf code to be run
on this metacomputer, we incorporated the message passing routines of the
Parallel Virtual Machine (PVM) software into the public-domain genetic
algorithm PIKAIA.

\subsection{Parallel Virtual Machine}

The PVM software \citep{gei94} allows a collection of networked computers
to cooperate on a problem as if they were a single multi-processor
parallel machine. All of the software and documentation was free. We had
no trouble installing it, and the sample programs that came with the
distribution made it easy to learn how to use.  The trickiest part of the
whole procedure was figuring out how to split up the workload among the
various computers.

The GA-based fitting procedure for the white dwarf code quite naturally
divided into two basic functions: evolving and pulsating white dwarf
models, and manipulating the results from each generation of trials. When
we profiled the distribution of execution time for each part of the code,
this division became even more obvious. The majority of the computing time
was spent evolving the starter model to a specific temperature. The GA is
concerned only with collecting and organizing the results of {\it many} of
these models, so it seemed reasonable to allocate many {\it slave}
computers to carry out the model calculations while a {\it master}
computer took care of the GA-related tasks.

In addition to decomposing the function of the code, a further division
based on the data was also possible. Since there were many trials in each
generation, the data required by the GA could easily be split into small,
computationally manageable units. One model could be sent to each
available slave computer, so the number of machines available would
control the number of models which could be calculated at the same time.

One minor caveat to the decomposition of the data into separate models to
be calculated by different computers is the fact that half of the machines
are slightly faster than the other half. Much of the potential increase in
efficiency from this parallelizing scheme could be lost if fast machines
are not sent more models to compute than slow ones. This may seem trivial,
but there is no mechanism built in to the current version of the PVM
software to handle this procedure automatically.

It is also potentially problematic to send out new jobs only after
receiving the results of previous jobs because the computers sometimes
hang or crash.  Again, this may seem obvious---but unless specifically
asked to check, PVM cannot tell the difference between a crashed computer
and one that simply takes a long time to compute a model. At the end of a
generation of trials, if the master process has not received the results
from one of the slave jobs, it would normally just continue to wait for 
the response indefinitely.

\subsection{The PIKAIA Subroutine}

PIKAIA is a self-contained, genetic-algorithm-based optimization
subroutine developed by Paul Charbonneau and Barry Knapp at the High
Altitude Observatory in Boulder, Colorado. Most optimization techniques
work to {\it minimize} a quantity---like the root-mean-square (r.m.s.)
residuals; but it is more natural for a genetic algorithm to {\it maximize} 
a quantity---natural selection works through survival of the
fittest. So PIKAIA maximizes a specified FORTRAN function through a call
in the body of the main program.

Unlike many GA packages available commercially or in the public domain,
PIKAIA uses decimal (rather than binary) encoding. Binary operations are
usually carried out through platform-dependent functions in FORTRAN, which
makes it more difficult to port the code between the Intel and Sun
platforms.

PIKAIA incorporates only the two basic genetic operators: uniform
one-point crossover, and uniform one-point mutation. The mutation rate can
be dynamically adjusted during the evolution, using either the linear
distance in parameter-space or the difference in fitness between the best
and median solutions in the population.  The practice of keeping the best
solution from each generation is called elitism, and is a default option
in PIKAIA.  Selection is based on ranking rather than absolute fitness,
and makes use of the Roulette Wheel algorithm. There are three different
reproduction plans available in PIKAIA: Steady-State-Delete-Random,
Steady-State-Delete-Worst, and Full Generational Replacement. Only the
last of these is easily parallelizable.

\vspace*{-6pt}
\section{Master Program}

\begin{figure}[p]
\hskip 0.5in
\epsfxsize 4.5in
\epsffile{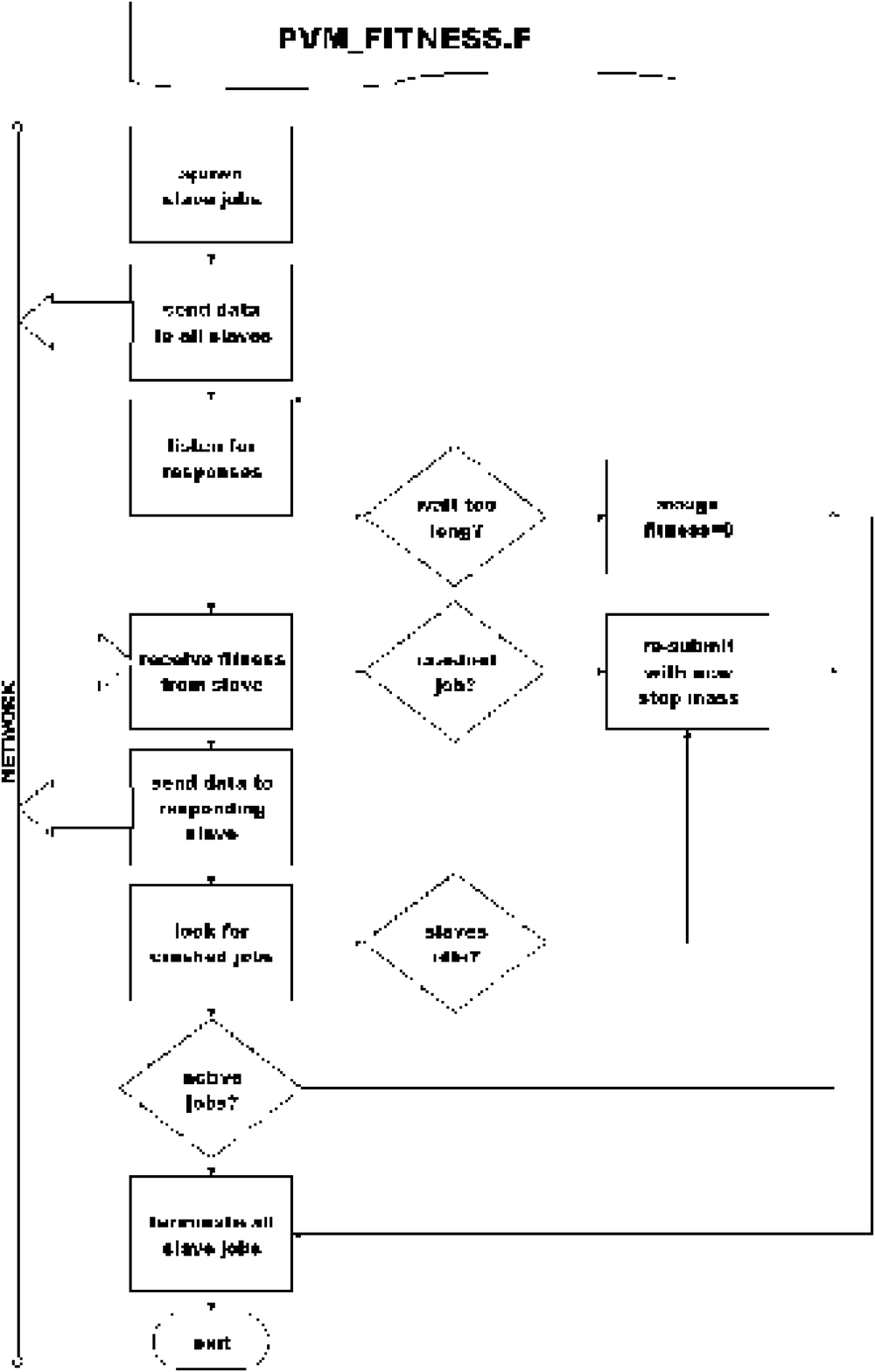}
\caption[Flow chart for the parallel fitness evaluation subroutine]
{Flow chart for the parallel fitness evaluation subroutine, which runs on
the master computer.\label{fig3.1}}
\end{figure}

\vspace*{-6pt}
Starting with an improved unreleased version of PIKAIA, we incorporated
the message passing routines of PVM into a parallel fitness evaluation
subroutine. The original code evaluated the fitnesses of the population of
trials one at a time in a DO loop. We replaced this procedure with a
single call to a new subroutine that evaluates the fitnesses in parallel
on all available processors.
{\footnotesize
\begin{verbatim}
c     initialize (random) phenotypes
      do ip=1,np
         do k=1,n
            oldph(k,ip)=urand()
         enddo
c     calculate fitesses
c        fitns(ip) = ff(n,oldph(1,ip))
      enddo
c     calculate fitnesses in parallel
      call pvm_fitness('ff_slave', np, n, oldph, fitns)
\end{verbatim}}
\noindent The parallel version of PIKAIA constitutes the master program
which runs on Darwin, the central computer in the network. A full listing
of the parallel fitness evaluation subroutine (PVM\_FITNESS.F) is included
in Appendix \ref{code}. A flow chart for this code is shown in Figure 
\ref{fig3.1}.

% FIGURE 3.1 %%%%%%%%%%%%%%%%%%%%%%%%%%%%%%%%%%%%%%%%%%%%%%%%%%%%%%%%%%%%%%
%%%%%%%%%%%%%%%%%%%%%%%%%%%%%%%%%%%%%%%%%%%%%%%%%%%%%%%%%%%%%%%%%%%%%%%%%%%

After starting the slave program on every available processor (64 for our
metacomputer), PVM\_FITNESS.F sends an array containing the values of the
parameters to each slave job over the network. In the first generation of
the GA, these values are completely random; in subsequent generations,
they are the result of the selection and mutation of the previous
generation, performed by the non-parallel portions of PIKAIA.

% FIGURE 3.2 %%%%%%%%%%%%%%%%%%%%%%%%%%%%%%%%%%%%%%%%%%%%%%%%%%%%%%%%%%%%%%
\begin{figure}[b]
\hskip 1.0in
\epsfxsize 3.5in
\epsffile{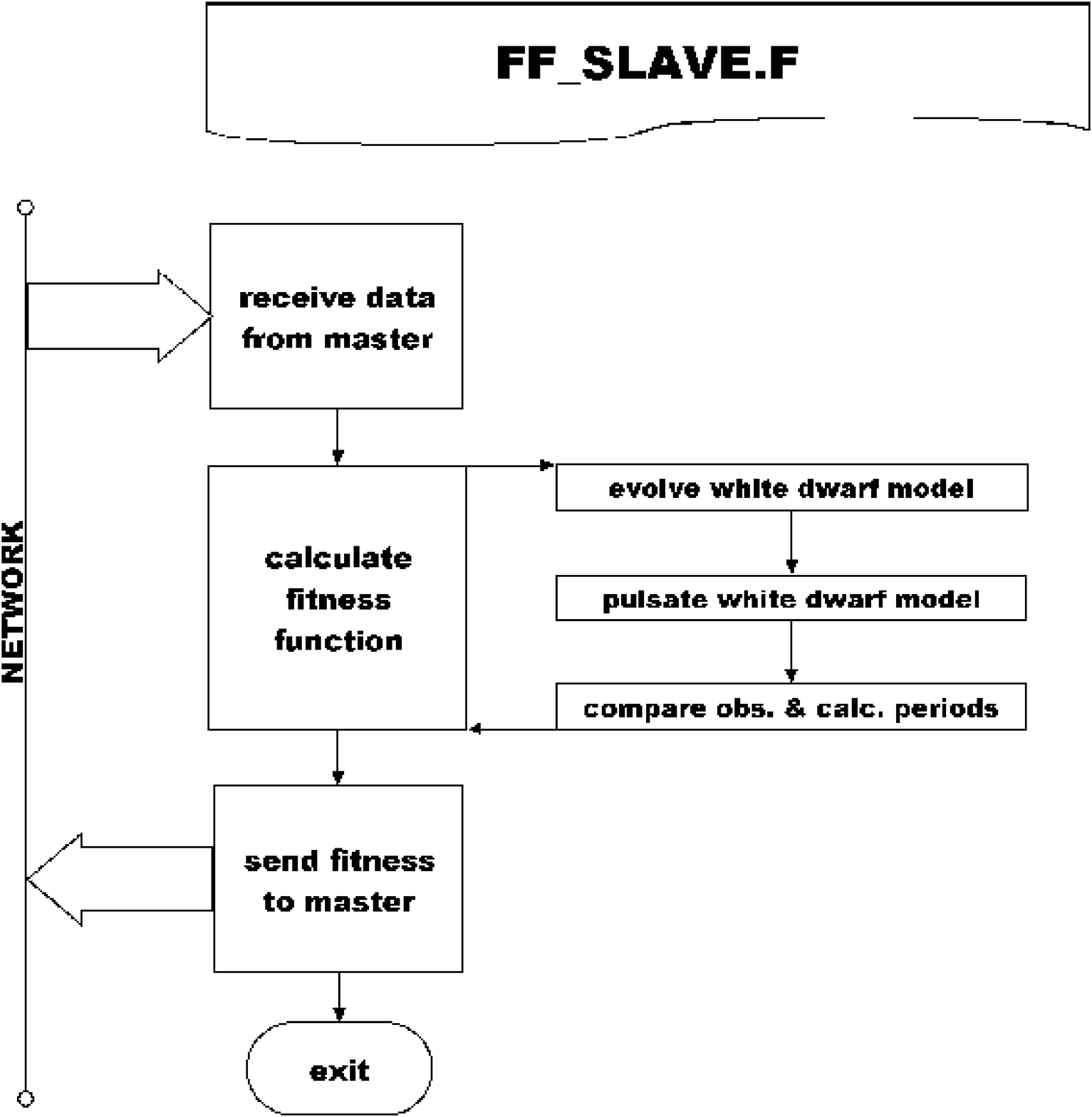}
\caption[Flow chart for the slave program]
{Flow chart for the slave program of the parallel code, which runs on each
of the 64 nodes of the metacomputer.\label{fig3.2}}
\end{figure}
%%%%%%%%%%%%%%%%%%%%%%%%%%%%%%%%%%%%%%%%%%%%%%%%%%%%%%%%%%%%%%%%%%%%%%%%%%%

Next, the subroutine listens for responses from the network and sends a
new set of parameters to each slave job as it finishes the previous
calculation.  When all sets of parameters have been sent out, the
subroutine begins looking for jobs that seem to have crashed and
re-submits them to slaves that have finished and would otherwise sit idle.
If a few jobs do not return a fitness after about five times the average
runtime required to compute a model, the subroutine assigns them a fitness
of zero. When every set of parameters in the generation have been assigned
a fitness value, the subroutine returns to the main program to perform the
genetic operations resulting in a new generation of models to calculate. The 
process continues for a fixed number of generations, chosen to maximize the 
efficiency of the search. The optimal number of generations is determined 
by applying the method to a test problem with a known solution.

\section{Slave Program}

The original white dwarf code came in three pieces: (1) the evolution
code, which evolves a starter model to a specific temperature, (2) the
prep code, which converts the output of the evolution code into a
different format, and (3) the pulsation code, which uses the output of the
prep code to determine the pulsation periods of the model.

To get the white dwarf code to run in an automated way, we merged the
three components of the original code into a single program, and added a
front end that communicated with the master program through PVM routines.
This code (FF\_SLAVE.F) constitutes the slave program, and is run on each
node of the metacomputer. A full listing of this code is included in
Appendix \ref{code}, and a flow chart is shown in Figure \ref{fig3.2}.

The operation of the slave program is relatively simple. Once it is
started by the master program, it receives a set of parameters from the
network. It then calls the fitness function (the white dwarf code) with
these parameters as arguments. The fitness function evolves a white dwarf
model with characteristics specified by the parameters, determines the
pulsation periods of this model, and then compares the calculated periods
to the observed periods of a real white dwarf. A fitness based on how well
the two sets of periods match is returned to the main program, which sends
it to the master program over the network. The node is then ready to run
the slave program again and receive a new set of parameters from the
master program.

\vfill\cleardoublepage\thispagestyle{empty}

\chapter{Forward Modeling \label{fwd}}

\begin{quote}
``Why do we always find a lost screwdriver in the last place we look?'' 

\hskip 2.5in{\it ---Joe \citeauthor{wam73}}
\end{quote}

\section{Introduction}

Having developed the hardware and software for the genetic-algorithm-based
approach to model fitting, we were finally ready to learn something about
white dwarf stars. There are presently three known classes of pulsating
white dwarfs. The hottest class are the planetary nebula nucleus variables
(PNNVs), which have atmospheres of ionized helium and are also called
DOVs. These objects require detailed calculations that evolve a main
sequence stellar model to the pre-white dwarf phase to yield accurate
pulsation periods. The two cooler classes are the helium-atmosphere
variable (DBV)  and hydrogen-atmosphere variable (DAV) white dwarfs. The
pulsation periods of these objects can be calculated accurately by
evolving simpler, less detailed models called polytropes. The DAV stars
are generally modeled as a core of carbon and oxygen with an overlying
blanket of helium covered by a thin layer of hydrogen on the surface. The
DBV stars are the simplest of all, with no detectable hydrogen---only a
helium layer surrounding the carbon/oxygen core. In the spirit of solving
the easier problem first, we decided to apply the GA method to the DBV
star GD~358.

\section{The DBV White Dwarf GD~358}

During a survey of eighty-six suspected white dwarf stars in the Lowell GD
lists, \cite{gre69} classified GD~358 as a helium atmosphere (DB) white
dwarf based on its spectrum. Photometric $UBV$ and $ubvy$ colors were
later determined by \cite{bw73} and \cite{weg79} respectively. Time-series
photometry by \cite{wrn82} revealed the star to be a pulsating
variable---the first confirmation of a new class of variable (DBV) white
dwarfs predicted by \cite{win81}.

In May 1990, GD~358 was the target of a coordinated observing run with the
Whole Earth Telescope \citep[WET;][]{nat90}. The results of these
observations were reported by \cite{win94}, and the theoretical
interpretation was given in a companion paper by \cite{bw94b}. They found
a series of nearly equally-spaced periods in the power spectrum which they
interpreted as non-radial {\it g}-mode pulsations of consecutive radial
overtone.  They attempted to match the observed periods and the period
spacing for these modes using earlier versions of the same theoretical
models we have used in this analysis (see \S \ref{modsect}). Their
optimization method involved computing a grid of models near a first guess
determined from general scaling arguments and analytical relations
developed by \cite{kaw90}, \cite{kw90}, \cite{bfwh92}, and \cite{bww93}.

\section{DBV White Dwarf Models \label{modsect}}

\subsection{Defining the Parameter-Space}

The most important parameters affecting the pulsation properties of DBV
white dwarf models are the total stellar mass ($M_*$), the effective
temperature ($T_{\rm eff}$), and the mass of the atmospheric helium layer
($M_{\rm He}$). We wanted to be careful to avoid introducing any
subjective bias into the best-fit determination simply by defining the
range of the search too narrowly. For this reason, we specified the range
for each parameter based only on the physics of the model, and on
observational constraints.

The distribution of masses for isolated white dwarf stars, generally
inferred from measurements of $\log\, g$, is strongly peaked near 0.6
\Msun\ with a full width at half maximum (FWHM)  of about 0.1 \Msun\
\citep{ngs99}. Isolated main sequence stars with masses near the limit for
helium ignition produce C/O cores more massive than about 0.45 \Msun, so
white dwarfs with masses below this limit must have helium cores
\citep{sgr90,ngs99}. However, the universe is not presently old enough to
produce helium core white dwarfs through single star evolution.  We
confine our search to masses between 0.45 \Msun\ and 0.95 \Msun. Although
some white dwarfs are known to be more massive than the upper limit of our
search, these represent a very small fraction of the total population and,
for reasonable assumptions about the mass-radius relation, all known DBVs
appear to have masses within the range of our search \citep{bea99}.

The span of temperatures within which DB white dwarfs are pulsationally
unstable is known as the DB instability strip.  The precise location of
this strip is the subject of some debate, primarily because of
difficulties in matching the temperature scales from ultraviolet and
optical spectroscopy and the possibility of hiding trace amounts of
hydrogen in the envelope \citep{bea99}. The most recent temperature
determinations for the 8 known DBV stars were done by \cite{bea99}. These
measurements, depending on various assumptions, place the red edge as low
as 21,800 K, and the blue edge as high as 27,800 K. Our search includes
all temperatures between 20,000 K and 30,000 K.

\begin{figure}
\hskip 0.5in
\epsfxsize 4.5in
\epsffile{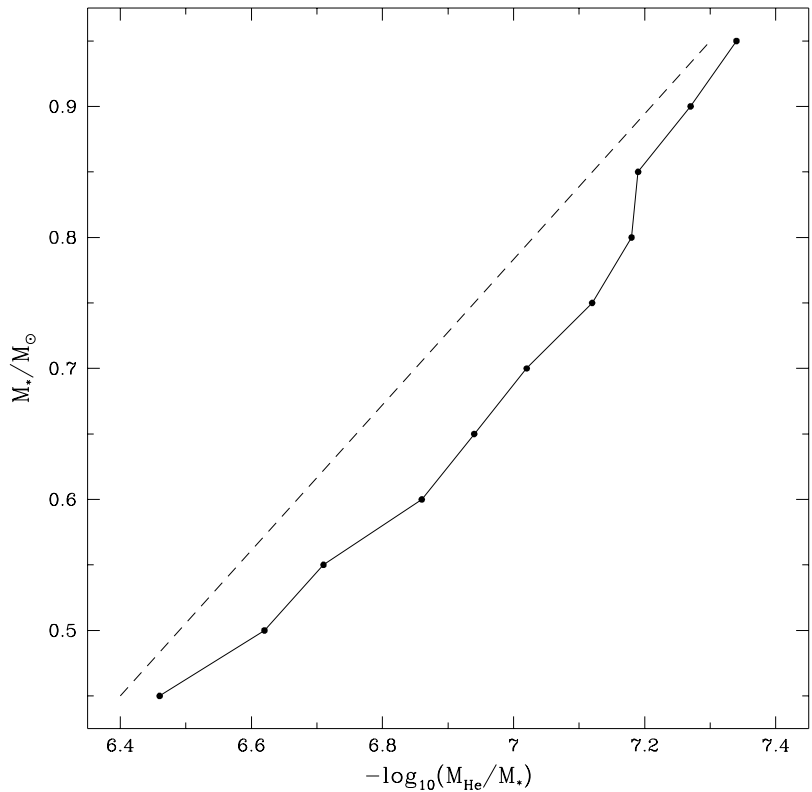}
\caption[Linear cut imposed on the helium layer mass]
{The numerical thin limit of the fractional helium layer mass for various
values of the total mass (connected points)  and the linear cut
implemented in the WD-40 code (dashed line).  \label{fig4.1}}
\end{figure}

The mass of the atmospheric helium layer must not be greater than about
$10^{-2}\ M_*$ or the pressure of the overlying material would
theoretically initiate helium burning at the base of the envelope. At the
other extreme, none of our models pulsate for helium layer masses less
than about $10^{-8}\ M_*$ over the entire temperature range we are
considering \citep{bw94a}. The practical limit is actually slightly larger
than this theoretical limit, and is a function of mass. For the most
massive white dwarfs we consider, our models run smoothly with a helium
layer as thin as $5\times 10^{-8}\ M_*$, while for the least massive the
limit is $4\times 10^{-7}\ M_*$ (see Figure \ref{fig4.1}).

% FIGURE 4.1 %%%%%%%%%%%%%%%%%%%%%%%%%%%%%%%%%%%%%%%%%%%%%%%%%%%%%%%%%%%%%%
%%%%%%%%%%%%%%%%%%%%%%%%%%%%%%%%%%%%%%%%%%%%%%%%%%%%%%%%%%%%%%%%%%%%%%%%%%%

\subsection{Theoretical Models}

To find the theoretical pulsation modes of a white dwarf, we start with a
static model of a pre-white dwarf and allow it to evolve quasi-statically
until it reaches the desired temperature. We then calculate the adiabatic
non-radial oscillation frequencies for the output model.  The initial
`starter' models can come from detailed calculations that evolve a
main-sequence star all the way to its pre-white dwarf phase, but this is
generally only important for accurate models of the hot DO white dwarfs.  
For the cooler DB and DA white dwarfs, it is sufficient to start with a
hot polytrope of order 2/3 (i.e. $P\propto\rho^{5/3}$).  The cooling
tracks of these polytropes converge with those of the pre-white dwarf
models well above the temperatures at which DB and DA white dwarfs are
observed to be pulsationally unstable \citep{woo90}.

To allow fitting for the total mass, we generated a grid of 100 starter
models with masses between 0.45 and 0.95 \Msun. The entire grid originated
from a 0.6~\Msun\ carbon-core polytrope starter model. We performed a
homology transform on this model to generate three new masses: 0.65, 0.75,
and 0.85 \Msun. We relaxed each of these three, and then used all four to
generate the grid. All models with masses below 0.6 \Msun\ were generated
by a direct homology transform of the original 0.6 \Msun\ polytrope. For
masses between $0.605 \rightarrow 0.745$ \Msun\ and from $0.755
\rightarrow 0.895$ \Msun, we used homology transforms of the relaxed 0.65
\Msun\ and 0.75 \Msun\ models respectively. The models with masses greater
than 0.9 \Msun\ were homology transforms of the relaxed 0.85 \Msun\ model.

To evolve a starter model to a specific temperature, we used the White
Dwarf Evolution Code (WDEC) described in detail by \cite{lv75} and by
\cite{woo90}. This code was originally written by Martin Schwarzschild,
and has subsequently been updated and modified by many others including:
\cite{ks69}, \cite{lv75}, \cite{win81}, \cite{kaw86}, \cite{woo90},
\cite{bra93}, and \cite{mon98}. The equation of state (EOS) for the cores
of our models come from \cite{lam74}, and from \cite{fgv77} for the
envelopes. We use the updated OPAL opacity tables from \cite{ir93},
neutrino rates from \cite{ito96}, and the ML3 mixing-length prescription
of \cite{bc71}. The evolution calculations for the core are fully
self-consistent, but the envelope is treated separately. The core and
envelope are stitched together and the envelope is adjusted to match the
boundary conditions at the interface. Adjusting the helium layer mass
involves stitching an envelope with the desired thickness onto the core
before starting the evolution. Because this is done while the model is
still very hot, there is plenty of time to reach equilibrium before the
model approaches the final temperature.

We determined the pulsation frequencies of the output models using the
adiabatic non-radial oscillation (ANRO) code described by \cite{kaw86},
originally written by Carl Hansen, which solves the pulsation equations
using the Runge-Kutta-Fehlberg method.

We have made extensive practical modifications to these programs,
primarily to allow models to be calculated without any intervention by the
user. The result is a combined evolution/pulsation code that runs smoothly
over a wide range of input parameters. We call this new code WD-40. Given
a mass, temperature, and helium layer mass within the ranges discussed
above, WD-40 will evolve and pulsate the specified white dwarf model and
return a list of the theoretical pulsation periods.

\section{Model Fitting}

Using the parallel version of PIKAIA on our metacomputer, we fixed the
population size at 128 trials, and initially allowed the GA to run for 250
generations. We used 2-digit decimal encoding for each of the three
parameters, which resulted in a temperature resolution of 100 K, a mass
resolution of 0.005 \Msun, and a resolution for the helium layer thickness
of 0.05 dex. The uniform single-point crossover probability was fixed at
85\%, and the mutation rate was allowed to vary between 0.1\% and 16.6\%,
depending on the linear distance in parameter-space between the trials
with the median and the best fitnesses.

\subsection{Application to Noiseless Simulated Data \label{inmodsect}}

To quantify the efficiency of our method for this problem, we used the
WD-40 code to calculate the pulsation periods of a model within the search
space, and then attempted to find the set of input parameters [$T_{\rm
eff}=25,000$~K, $M_* = 0.600$ \Msun, $\log (M_{\rm He}/M_*) = -5.96$]
using the GA. We performed 20 independent runs using different random
initialization each time. The first order solutions found in each case by
the GA are listed in Table \ref{tab4.1}. In 9 of the 20 runs, the GA found
the exact set of input parameters, and in 4 other runs it finished in a
region of parameter-space close enough for a small (1331 point) grid to
reveal the exact answer. Since none of the successful runs converged
between generations 200 and 250, we stopped future runs after 200
generations.

% TABLE 4.1 %%%%%%%%%%%%%%%%%%%%%%%%%%%%%%%%%%%%%%%%%%%%%%%%%%%%%%%%%%%%%%%
\begin{table}
\begin{center}
\caption{\label{tab4.1}Results for Noiseless Simulated Data}
\vskip 5pt
\begin{tabular}{cccccc}
\hline\hline
& \multicolumn{3}{c}{First-Order Solution} & & Generation \\ \cline{2-4}
Run & $T_{\rm eff}$ & $M_*/M_{\odot}$ & $\log(M_{\rm He}/M_*)$ & r.m.s.& Found \\
\hline
01 & 26,800 & 0.560 & $-$5.70 & 0.67 & 245 \\
02 & 25,000 & 0.600 & $-$5.96 & 0.00 & 159 \\
03 & 24,800 & 0.605 & $-$5.96 & 0.52 & 145 \\
04 & 25,000 & 0.600 & $-$5.96 & 0.00 &  68 \\
05 & 22,500 & 0.660 & $-$6.33 & 1.11 &  97 \\
06 & 25,000 & 0.600 & $-$5.96 & 0.00 & 142 \\
07 & 25,000 & 0.600 & $-$5.96 & 0.00 &  97 \\
08 & 25,000 & 0.600 & $-$5.96 & 0.00 & 194 \\
09 & 25,200 & 0.595 & $-$5.91 & 0.42 & 116 \\
10 & 26,100 & 0.575 & $-$5.80 & 0.54 &  87 \\
11 & 23,900 & 0.625 & $-$6.12 & 0.79 &  79 \\
12 & 25,000 & 0.600 & $-$5.96 & 0.00 & 165 \\
13 & 26,100 & 0.575 & $-$5.80 & 0.54 &  92 \\
14 & 25,000 & 0.600 & $-$5.96 & 0.00 &  95 \\
15 & 24,800 & 0.605 & $-$5.96 & 0.52 &  42 \\
16 & 26,600 & 0.565 & $-$5.70 & 0.72 & 246 \\
17 & 24,800 & 0.605 & $-$5.96 & 0.52 & 180 \\
18 & 25,000 & 0.600 & $-$5.96 & 0.00 &  62 \\
19 & 24,100 & 0.620 & $-$6.07 & 0.76 & 228 \\
20 & 25,000 & 0.600 & $-$5.96 & 0.00 & 167 \\
\hline\hline
\end{tabular}
\end{center}
\vskip -12pt
\end{table}
%%%%%%%%%%%%%%%%%%%%%%%%%%%%%%%%%%%%%%%%%%%%%%%%%%%%%%%%%%%%%%%%%%%%%%%%%%%

From the 13 runs that converged in 200 generations, we deduce an
efficiency for the method (GA + small grid) of $\sim$65\%. This implies
that the probability of missing the correct answer in a single run is
$\sim$35\%.  By running the GA several times, we reduce the probability of
not finding the correct answer: the probability that two runs will both be
wrong is $\sim$12\%, for three runs it is $\sim$4\%, and so on. Thus, to
reduce the probability of not finding the correct answer to below 1\% we
need to run the GA, on average, 5 times. For 200 generations of 128
trials, this requires $\sim$10$^5$ model evaluations. By comparison, an
exhaustive search of the parameter-space with the same resolution would
require $10^6$ model evaluations, so our method is comparably global but
about 10$\times$ more efficient than an exhaustive search of
parameter-space. Even with this efficiency and our ability to run the
models in parallel, each run of the GA required about 6 hours to complete.

\subsection{The Effect of Gaussian Noise \label{noisesect}}

Having established that the GA could find the correct answer for noiseless
data, we wanted to see how noise on the frequencies might affect it.
Before adding noise to the input frequencies, we attempted to characterize
the actual noise present on frequencies determined from a WET campaign. We
approached this problem in two different ways.

First, we tried to characterize the noise empirically by looking at the
differences between the observed and predicted linear combination
frequencies. When we take the Fourier Transform of the long light curves
from WET observations, we are effectively decomposing the signal into
perfect sinusoidal components of various frequencies. The actual light
variations arising from a single pulsation mode are generally not perfect
sinusoids, and the amplitude often varies on relatively short timescales.
This leads to significant power in the Fourier Transform at integer
multiples and fractions of the real pulsation frequency. These so-called
harmonics and sub-harmonics are examples of linear combination
frequencies, but they are not nearly as prevalent as the combinations that
arise from the interaction of different pulsation modes in the observed
light curves.

When two pulsation modes with different frequencies $f_1$ and $f_2$ are
present, the resulting light curve shows oscillations at the sum of the
two frequencies $f_1+f_2$. As the frequencies go into and out of phase or
``beat'' with each other, the amplitude of the light variation grows and
shrinks in a periodic manner at a frequency equal to the difference of the
two components $f_1-f_2$. If the amplitude of the beating is smaller than
expected due to some non-linear effect, then the Fourier Transform will
reveal power not only at the real pulsation frequencies, but also at these
combination frequencies. If we can properly identify the pulsation modes
that produce a specific linear combination, it allows us to estimate the
uncertainty on our frequency measurements because the combination should
show up at a precise sum or difference.

\begin{figure}[b]
\hskip 0.75in
\epsfxsize 4.0in
\epsffile{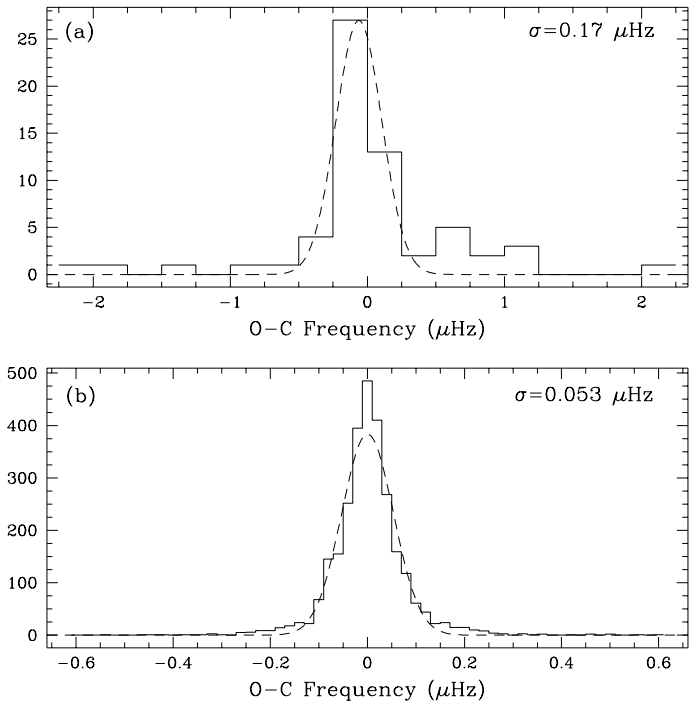}
\caption[Two determinations of the noise on pulsation frequencies]
{The distribution of differences between (a) the observed and predicted
frequencies of linear combination modes identified by \cite{vui00} in the
1994 WET run on GD~358 and the best-fit Gaussian with $\sigma=0.17\ \mu$Hz
(dashed line) and (b) the input and output frequencies for the 11 modes
used to model GD~358 from simulated WET runs (see \S \ref{noisesect} for
details) and the best-fit Gaussian with $\sigma=0.053\ \mu$Hz (dashed
line).\label{fig4.2}}
\end{figure}

We used the mode identifications of \cite{vui00} from the 1994 WET run on
GD~358. There were a total of 63 combinations identified: 20 sum and 11
difference frequencies of 2-mode combinations, 30 sum and difference
3-mode combinations, and 2 combinations involving 4 modes. We used the
measured frequencies of the parent modes to predict the frequency of each
linear combination, and then compared this to the observed frequency.  
The distribution of observed minus computed frequencies for these 63
modes, and the best-fit Gaussian is shown in the top panel of Figure
\ref{fig4.2}. The distribution has $\sigma=0.17\ \mu$Hz.

% FIGURE 4.2 %%%%%%%%%%%%%%%%%%%%%%%%%%%%%%%%%%%%%%%%%%%%%%%%%%%%%%%%%%%%%%
%%%%%%%%%%%%%%%%%%%%%%%%%%%%%%%%%%%%%%%%%%%%%%%%%%%%%%%%%%%%%%%%%%%%%%%%%%%

Second, we tried to characterize the noise by performing the standard
analysis for WET runs on many simulated light curves to look at the
distribution of differences between the input and output frequencies. We
generated 100 synthetic GD~358 light curves using the 57 observed
frequencies and amplitudes from \cite{win94}. Each light curve had the
same time span as the 1990 WET run (965,060 seconds) sampled with the same
interval (every 10 seconds) but without any gaps in coverage. Although the
noise in the observed light curves was clearly time-correlated, we found
that the distribution around the mean light level for the comparison star
after the standard reduction procedure was well represented by a Gaussian.  
So we added Gaussian noise to the simulated light curves to yield a
signal-to-noise ratio $S/N \approx 2$, which is typical of the observed
data. We took the discrete Fourier Transform of each light curve, and
identified peaks in the same way as is done for real WET runs. We
calculated the differences between the input frequencies and those
determined from the simulation for the 11 modes used in the seismological
analysis by \cite{bw94b}. The distribution of these differences is shown
in the bottom panel of Figure \ref{fig4.2}, along with the best-fit
Gaussian which has $\sigma=0.053\ \mu$Hz. Because the noise measurement
from linear combination frequencies suffered from low-number statistics,
we adopted the noise estimate from the synthetic light curves for our 
studies of the effect of noise on the GA method.

Using the same input model as in \S \ref{inmodsect}, we added random
offsets drawn from a Gaussian distribution with $\sigma=0.053\ \mu$Hz to
each of the frequencies.  We produced 10 sets of target frequencies from
10 unique realizations of the noise, and then applied the GA method to
each set.  In all cases the best of 5 runs of the GA found the exact set
of parameters from the original input model, or a region close enough to
reveal the exact solution after calculating the small grid around the
first guess.  To reassure ourselves that the success of the GA did not
depend strongly on the amount of noise added to the frequencies, we also
performed fits for several realizations of the larger noise estimate from
the analysis of linear combination frequencies. The method always
succeeded in finding the original input model.

\vspace*{-8pt}
\subsection{Application to GD~358}
\vspace*{-4pt}

Having thoroughly characterized the GA method, we finally applied it to
real data. We used the same 11 periods used by \cite{bw94b}. As in their
analysis, we assumed that the periods were consecutive radial overtones
and that they were all $\ell=1$ modes 
\citep[][give detailed arguments to support this assumption]{win94}. 
Anticipating that the GA might have
more difficulty with non-synthetic data, we decided to perform a total of
10 GA runs for each core composition. This should reduce the chances of
not finding the best answer to less than about 3 in 10,000.

To facilitate comparison with previous results, we obtained fits for six
different combinations of core composition and internal chemical profiles:
pure C, pure O, and both ``steep'' and ``shallow'' internal chemical
profiles for 50:50 and 20:80 C/O cores \citep[see][]{bww93,bw94b}.

We also ran the GA with an alternate fitness criterion for the 50:50 C/O
``steep'' case, which contains the best-fit model of \cite{bw94b}.
Normally, the GA only attempts to match the pulsation periods. We
reprogrammed it to match both the periods and the period spacing, which
was the fitness criterion used by \citeauthor{bw94b}. Within the range of
parameters they considered, using this alternate fitness criterion, the GA
found best-fit model parameters consistent with \citeauthor{bw94b}'s
solution.

\vspace*{-16pt}
\section{Initial Results}
\vspace*{-8pt}

\begin{figure}
\hskip 0.25in
\epsfxsize 5.0in
\epsffile{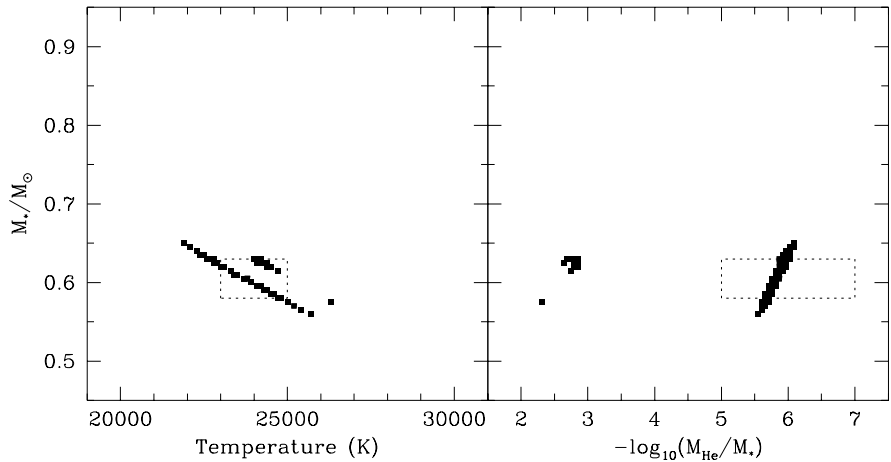}
\caption[General features of the 3-dimensional parameter-space]
{Front and side views of the GA search space for a C/O 50:50 core with a
``steep'' internal chemical profile. Square points mark the locations of every
model found by the GA with an r.m.s. deviation smaller than 3 seconds for
the periods observed in GD~358. The dashed line shows the range of
parameters considered by \cite{bw94b}. \label{fig4.3}}
\vspace*{-12pt}
\end{figure}

The general features of the 3-dimensional parameter-space for GD~358 are
illustrated in Figure \ref{fig4.3}.  All combinations of parameters found
by the GA for a 50:50 C/O steep core having r.m.s.~period differences
smaller than 3 seconds are shown as square points in this plot. The two
panels are orthogonal projections of the search space, so each point in
the left panel corresponds one-to-one with a point in the right panel.
Essentially, Figure \ref{fig4.3} shows which combinations of model
parameters yield reasonably good matches to the periods observed in GD~358
for this core composition.

The most obvious feature of the parameter-space is the presence of more
than one region that yields a good match to the observations. Generally,
the good fits seem to cluster in two families corresponding to thick and
thin helium layers. This is the first exciting result of applying the GA
method. The best-fit solution of \cite{bw94b} falls in the family with
thin helium layers. This solution was problematic at the time because the
earlier asteroseismological investigation of the hot white dwarf
PG~1159-035 \citep{win91} implied that it had a relatively thick helium
layer. If there is an evolutionary connection between PG~1159 stars and
the DBVs, it was more difficult to understand if the helium layers were
significantly different. If a better solution for GD~358 exists among the
family with thick helium layers, this long-standing controversy might be
resolved.

% FIGURE 4.3 %%%%%%%%%%%%%%%%%%%%%%%%%%%%%%%%%%%%%%%%%%%%%%%%%%%%%%%%%%%%%%
%%%%%%%%%%%%%%%%%%%%%%%%%%%%%%%%%%%%%%%%%%%%%%%%%%%%%%%%%%%%%%%%%%%%%%%%%%%

The other obvious features in Figure \ref{fig4.3} are the
parameter-correlations in both projections, causing the good fits to fall
along lines in parameter-space rather than on a single point. The
correlation between total mass and fractional helium layer mass
is relatively easy to understand. \cite{bfwh92} showed that the pulsation
periods of trapped modes in white dwarf models are strongly influenced by
the scaled location of the composition transition zone. They developed an
expression showing that these periods are directly proportional to the
{\it fractional} radius of the composition interface.  As the total mass
of a white dwarf increases, the surface area decreases, so the mass of
helium required to keep the interface at the same {\it fractional} radius
also decreases. Thus, a thinner helium layer can compensate for an
overestimate of the mass.

The correlation between mass and temperature is slightly more
complicated. The natural frequency that dominates the determination of
white dwarf model pulsation frequencies is the Brunt-V\"ais\"al\"a
frequency (named after meteorologists David Brunt and Yuri V\"ais\"al\"a)
which reflects the difference between the actual and the
adiabatic density gradients. As the temperature decreases, the matter
becomes more degenerate, so the Brunt-V\"ais\"al\"a frequency in much of
the star tends to zero. The pulsation periods of a white dwarf model in
some sense reflect the average of the Brunt-V\"ais\"al\"a frequency
throughout the star, so a decrease in temperature leads to lower pulsation
frequencies. Higher mass models have higher densities, which generally
lead to higher pulsation frequencies. So an overestimate of the mass can
compensate for the effect of an underestimate of the temperature.

% FIGURE 4.4 %%%%%%%%%%%%%%%%%%%%%%%%%%%%%%%%%%%%%%%%%%%%%%%%%%%%%%%%%%%%%%
\begin{figure}[b]
\hskip 0.25in
\epsfxsize 5.0in
\epsffile{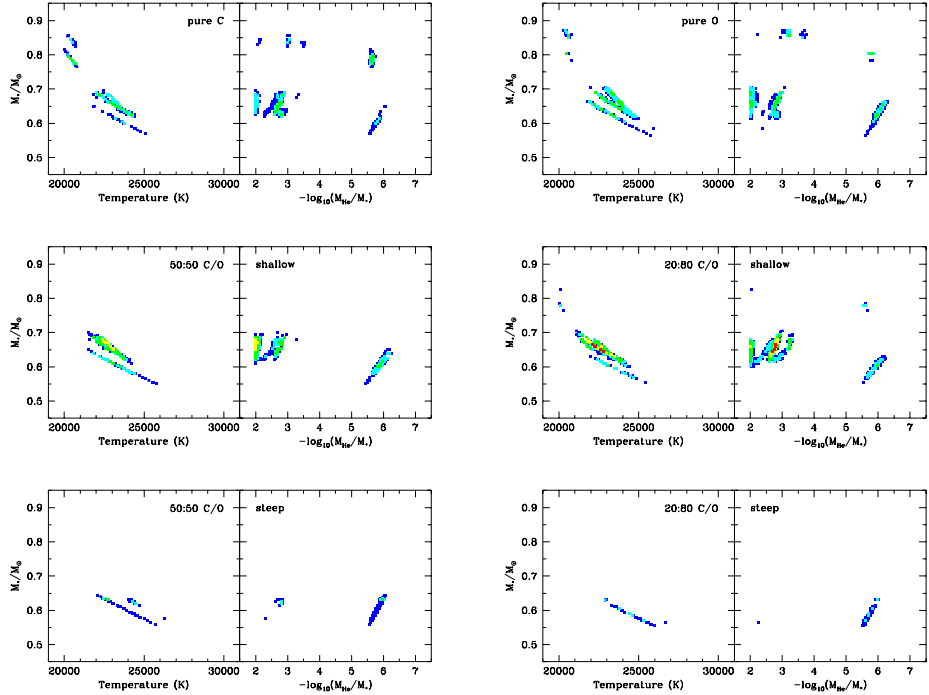}
\caption[Parameter-space for GD~358 using various core compositions]
{The families of models found by the GA which yield good matches to the
periods observed in GD~358 for various core compositions and internal chemical
profiles. The quality of the fit is indicated by the color of the square:
${\rm r.m.s.}<3.0$ seconds (blue), $<2.75$ seconds (cyan), $<2.5$ seconds
(green), $<2.0$ seconds (yellow), and $<1.65$ seconds (red).
\label{fig4.4}}
\end{figure}
%%%%%%%%%%%%%%%%%%%%%%%%%%%%%%%%%%%%%%%%%%%%%%%%%%%%%%%%%%%%%%%%%%%%%%%%%%%

The results for all six core compositions and internal chemical profiles
are shown in Figure \ref{fig4.4}, where we have used color to indicate the
absolute quality of each fit. We find that reasonable fits are possible
with every core composition, but excellent fits (indicated by red points
in the figure) are only possible for a specific core composition and
internal chemical profile. Pure C and pure O models appear to have more
families of possible solutions, but the high-mass families have
luminosities that can be ruled out based on the observed parallax of
GD~358 \citep{har85}.  Mixed C/O cores generally seem to produce better
fits, but internal chemical profiles that are steep are much worse than
those that are shallow. Among the mixed C/O cores with shallow internal
chemical profiles, the 20:80 mix produces the best fits of all.

% TABLE 4.2 %%%%%%%%%%%%%%%%%%%%%%%%%%%%%%%%%%%%%%%%%%%%%%%%%%%%%%%%%%%%%%%
\begin{table}[b]
\begin{center}
\caption{\label{tab4.2}Results for GD~358 Data}
\vskip 5pt
\begin{tabular}{lcccl}
\hline\hline
 & \multicolumn{3}{c}{Best Models} & \\ \cline{2-4}
Core Properties & $T_{\rm eff}$ & $M/M_{\odot}$& $\log(M_{\rm He}/M_*)$ & r.m.s.\\
\hline
pure C \dotfill                       & 20,300 & 0.795 & $-$5.66 & 2.17$^a$\\
                                      & 23,100 & 0.655 & $-$2.74 & 2.30 \\
 & & & & \\
50:50 C/O ``shallow''$\ldots$\dotfill & 22,800 & 0.665 & $-$2.00 & 1.76 \\
                                      & 23,100 & 0.610 & $-$5.92 & 2.46 \\
 & & & & \\
50:50 C/O ``steep''\dotfill           & 22,700 & 0.630 & $-$5.97 & 2.42 \\
                                      & 24,300 & 0.625 & $-$2.79 & 2.71 \\
 & & & & \\
20:80 C/O ``shallow''$\ldots$\dotfill & 22,600 & 0.650 & $-$2.74 & 1.50$^b$ \\
                                      & 23,100 & 0.605 & $-$5.97 & 2.48 \\
 & & & & \\
20:80 C/O ``steep'' \dotfill          & 22,900 & 0.630 & $-$5.97 & 2.69 \\
                                      & 27,300 & 0.545 & $-$2.16 & 2.87$^a$ \\
 & & & & \\
pure O \dotfill                       & 20,500 & 0.805 & $-$5.76 & 2.14$^a$ \\
                                      & 23,400 & 0.655 & $-$2.79 & 2.31 \\
\hline\hline
\end{tabular}
\end{center}
\vskip -6pt
\hskip 0.5in {$^a$ Luminosity is inconsistent with observations}

\hskip 0.5in {$^b$ Best-fit solution}
\end{table}
%%%%%%%%%%%%%%%%%%%%%%%%%%%%%%%%%%%%%%%%%%%%%%%%%%%%%%%%%%%%%%%%%%%%%%%%%%%

The parameters for the best-fit models and measures of their absolute
quality are listed in Table \ref{tab4.2}. For each core composition, the
best-fit for both the thick and thin helium layer families are shown. As
indicated, several fits can be ruled out based on their luminosities. Our
new best-fit model for GD~358 has a mass and temperature marginally
consistent with those inferred from spectroscopy.

\section{Internal Composition \& Structure \label{FWDSEC}}

The initial 3-parameter fits to GD~358 make it clear that both the central
oxygen abundance and the shape of the internal chemical profile should be
treated as free parameters. We modified our code to allow any central
oxygen mass fraction ($X{\rm _O}$) between 0.00 and 1.00 with a resolution
of 1 percent. To explore different chemical profiles we fixed $X{\rm _O}$
to its central value out to a fractional mass parameter ($q$) which varied
between 0.10 and 0.85 with a resolution of 0.75 percent. From this point,
we forced $X{\rm _O}$ to decrease linearly in mass to zero oxygen at the
95 percent mass point.

This parameterization is a generalized form of the ``steep'' and
``shallow'' profiles. We used these profiles so that our results could be
easily compared to earlier work by \cite{woo90} and \cite{bww93}. The
latter authors define both profiles in their Figure 1. The ``shallow''
profile corresponds approximately to $q=0.5$, and ``steep'' corresponds
roughly to $q=0.8$. However, in our generalized parameterization we have
moved the point where the oxygen abundance goes to zero from a fractional
mass of 0.9 out to a fractional mass of 0.95. This coincides with the
boundary in our models between the self-consistent core and the envelope,
where we describe the He/C transition using a diffusion equilibrium
profile from the method of \cite{af80} with diffusion exponents of $\pm3$.
We do not presently include oxygen in the envelopes, so the mass fraction
of oxygen must drop to zero by this point.

We calculated the magnitude of deviations from the mean period spacing for
models using our profiles compared to those due to smooth profiles from
recent theoretical calculations by \cite{sal97}. The smooth theoretical
profiles caused significantly larger deviations, so we conclude that the
abrupt changes in the oxygen abundance resulting from our parameterization
do not have an unusually large effect on the period spacing. Although the
actual chemical profiles will almost certainly differ from the profiles
resulting from our simple parameterization, we should still be able to
probe the gross features of the interior structure by matching one or
another linear region of the presumably more complicated physical
profiles.

We used the same ranges and resolution for $M_{*}$, $T_{\rm eff}$, and
$M_{\rm He}$ as in the initial study, so the search space for this
5-parameter problem is 10,000 times larger than for the 3-parameter case.
Initially we tried to vary all 5 parameters simultaneously, but this
proved to be impractical because of the parameter-correlation between
$M_{*}$ and $q$. The pulsation properties of the models depend on the {\it
radial} location in the chemical profile where the oxygen mass fraction
begins to change. If the GA finds a combination of $M_{*}$ and $q$ that
yields a reasonably good fit to the data, most changes to either one of
them by itself will not improve the fit. As a consequence, simultaneous
changes to both parameters are required to find a better fit, and since
this is not very probable the GA must run for a very long time.  Tests on
synthetic data for the full 5-parameter problem yielded only a 10 percent
probability of finding the input model even when we ran for 2000
generations---ten times longer than for the 3-parameter case. By contrast,
when we used a fixed value of $q$ and repeated the test with only 4 free
parameters, the GA found the input model in only 400 generations for 8 out
of 10 runs.  Even better, by fixing the mass and allowing $q$ to vary, it
took only 250 generations to find the input model in 7 out of 10 runs.
This suggests that it might be more efficient to alternate between these
two subsets of 4 parameters, fixing the fifth parameter each time to its
best-fit value from the previous iteration, until the results of both fits
are identical.

Since we do not know {\it a priori} the precise mass of the white dwarf,
we need to ensure that this iterative 4-parameter approach will work even
when the mass is initially fixed at an incorrect value. To test this, we
calculated the pulsation periods of the best-fit $0.65\ M_{\odot }$ model
for GD~358 from Table \ref{tab4.2} and then iteratively applied the two
4-parameter fitting routines, starting with the mass fixed at $0.60\
M_{\odot }$---an offset comparable to the discrepancy between the mass
found in Table \ref{tab4.2} and the value found by \cite{bw94b}. The
series of fits leading to the input model are shown in Table \ref{tab4.3}.
This method required only 3 iterations, and for each iteration we
performed 10 runs with different random initialization to yield a
probability of finding the best-fit much greater than 99.9 percent. In the
end, the method required a total of $2.5\times 10^6$ model evaluations
(128 trials per generation, 10 runs of 650 generations per iteration).
This is about 200 times more efficient than calculating the full grid in
each iteration, and about 4,000 times more efficient than a grid of the
entire 5-dimensional space.

% TABLE 4.3 %%%%%%%%%%%%%%%%%%%%%%%%%%%%%%%%%%%%%%%%%%%%%%%%%%%%%%%%%%%%%%%
\begin{table}
\begin{center}
\caption{Convergence of the Method on Simulated Data\label{tab4.3}}
\vskip 5pt
\begin{tabular}{cccccc}
\hline\hline
Iteration & $T_{\rm eff}$ & $M_*/M_{\odot}$ & $\log(M_{\rm He}/M_*)$ & $X_{\rm O}$ & $q$ \\ 
\hline
1$^a$ & 23,600 & 0.600 & $-$5.76 & 0.52 & 0.55 \\
1$^b$ & 22,200 & 0.660 & $-$2.79 & 0.99 & 0.55 \\
2$^a$ & 22,200 & 0.660 & $-$2.79 & 0.88 & 0.51 \\
2$^b$ & 22,600 & 0.650 & $-$2.74 & 0.85 & 0.51 \\
3$^a$ & 22,600 & 0.650 & $-$2.74 & 0.80 & 0.50 \\
3$^b$ & 22,600 & 0.650 & $-$2.74 & 0.80 & 0.50 \\
\hline\hline
\end{tabular}
\end{center}
\vskip -0.1in
\hskip 1.1in{$^a$ Value of $M_*/M_{\odot}$ fixed}

\hskip 1.1in{$^b$ Value of $q$ fixed.}
\end{table}
%%%%%%%%%%%%%%%%%%%%%%%%%%%%%%%%%%%%%%%%%%%%%%%%%%%%%%%%%%%%%%%%%%%%%%%%%%%

Next, we applied this iterative 4-parameter method to the observed
pulsation periods of GD~358. We initially fixed the mass at $0.61\
M_{\odot}$, the value inferred from the original asteroseismological study
by \cite{bw94b}. The solution converged after four iterations, and the
best-fit values of the five parameters were:
\begin{flushleft}
\begin{tabular}{lrclcrcl}
~~~~~~~~~~~~~&$T_{\rm eff}$          &=& 22,600\ K & & $X_{\rm O}$ &=& 0.84 \\
~~~~~~~~~~~~~&$M_*/M_{\odot}$        &=& 0.650     & & $q$         &=& 0.49 \\
~~~~~~~~~~~~~&$\log(M_{\rm He}/M_*)$ &=& $-$2.74   & &             & &      \\
\end{tabular}
\end{flushleft}
Note that the values of $M_*$, $T_{\rm eff}$ and $M_{\rm He}$ are
identical to the best-fit in Table \ref{tab4.2}. The best-fit mass and
temperature still differ significantly from the values inferred from
spectroscopy by \cite{bea99}. However, the luminosity of our best-fit
model is consistent with the luminosity derived from the measured parallax
of GD~358 \citep{har85}.

\begin{figure}
\hskip 0.75in
\epsfxsize 4.0in
\epsffile{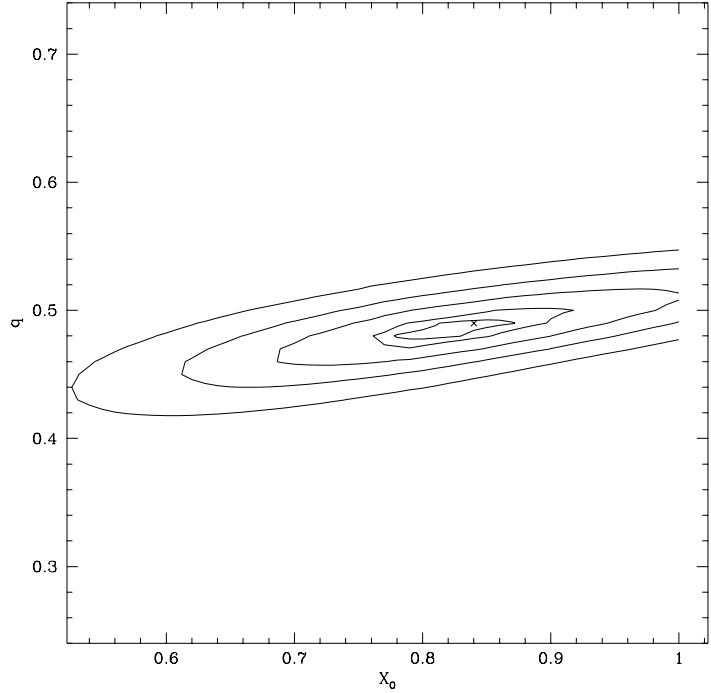}
\caption[Uncertainty on the central oxygen mass fraction in GD~358]
{A contour plot of the central oxygen mass fraction ($X_{\rm O}$) versus
the fractional mass location of the change in the oxygen gradient ($q$),
with $M_*$, $T_{\rm eff}$, and $M_{\rm He}/M_*$ fixed at their best-fit
values. The model with the absolute minimum residuals (identical to the
best-fit found by the GA) is marked with an X.  The contours are drawn at
1, 3, 10, 25 and 40 times the observational noise. \label{fig4.5}}
\end{figure}

To alleviate any doubt that the GA had found the best combination of
$X_{\rm O}$ and $q$, and to obtain a more accurate estimate of the
uncertainties on these parameters, we calculated a grid of 10,000 models
with the mass, temperature, and helium layer mass fixed at their best-fit
values. A contour plot of this grid near the solution found by the GA is
shown in Figure \ref{fig4.5}.

% FIGURE 4.5 %%%%%%%%%%%%%%%%%%%%%%%%%%%%%%%%%%%%%%%%%%%%%%%%%%%%%%%%%%%%%%%%
%%%%%%%%%%%%%%%%%%%%%%%%%%%%%%%%%%%%%%%%%%%%%%%%%%%%%%%%%%%%%%%%%%%%%%%%%%%

\section{Constraints on Nuclear Physics}

During carbon burning in a red giant star, the triple-$\alpha$ process
competes with the $^{12}{\rm C}(\alpha,\gamma)^{16}{\rm O}$ reaction for
the available $\alpha$-particles. As a consequence, the final ratio of
carbon to oxygen in the core of a white dwarf is a measure of the relative
cross-sections of these two reactions \citep{buc96}. The value of the
$^{12}$C$(\alpha ,\gamma )^{16}$O cross-section at stellar energies is
presently the result of an extrapolation across eight orders of magnitude
from laboratory data \citep{fow86}. The triple-$\alpha$ reaction is
relatively well determined at the relevant energies, so the constraint
that comes from measuring the C/O ratio in the core of a white dwarf is
much more precise than other methods.

Adopting the central oxygen mass fraction from the 5-parameter best-fit
forward modeling ($X_{\rm O}=84\pm3$ percent) we can place preliminary
constraints on the $^{12}$C$(\alpha ,\gamma )^{16}$O cross-section.
\cite{sal97} made detailed evolutionary calculations for main-sequence
stellar models with masses between 1 and 7 M$_{\odot}$ to provide internal
chemical profiles for the resulting white dwarfs. For the bulk of the
calculations they adopted the rate of \cite{cau85} for the
$^{12}$C$(\alpha ,\gamma )^{16}$O reaction ($S_{300}=240$ keV barns), but
they also computed an evolutionary sequence using the lower cross-section
inferred by \cite{wtw93} from solar abundances ($S_{300}=170$ keV barns).
The chemical profiles from both rates had the same general shape, but the
oxygen abundances were uniformly smaller for the lower rate. In both cases
the C/O ratio was constant out to the 50 percent mass point, a region
easily probed by white dwarf pulsations.

The central oxygen mass fraction is lower in higher mass white dwarf
models. The rate of the triple-$\alpha$ reaction (a three-body process)  
increases faster at higher densities than does the $^{12}$C$(\alpha
,\gamma )^{16}$O reaction. As a consequence, more helium is used up in the
production of carbon, and relatively less is available to produce oxygen
in higher mass models. Interpolating between the models of \cite{sal97}
which used the higher value of the cross-section, we expect a central
oxygen mass fraction for a $M_*=0.65\ M_{\odot}$ model of $X_{\rm O}^{\rm
high}=0.75$. Using additional calculations for the low rate (M.~Salaris
2001, private communication), the expected value is $X_{\rm O}^{\rm
low}=0.62$. Extrapolating to the value inferred from our 5-parameter
forward modeling, we estimate that the astrophysical S-factor at 300 keV
for the $^{12}$C$(\alpha , \gamma )^{16}$O cross-section is in the range
$S_{300}=290 \pm 15$ keV barns (internal uncertainty only).

\vfill\cleardoublepage\thispagestyle{empty}

\chapter{Reverse Approach \label{rev}}

\begin{quote}
``The only real voyage of discovery consists not in seeking 
new landscapes but in having new eyes.'' 

\hskip 2.5in{\it ---Marcel \citeauthor{pro81}}
\end{quote}

\section{Introduction}

The results of forward modeling with one adjustable point in the chemical
profile make it clear that information about the internal structure is
contained in the data, and we just need to learn how to extract it. If we
want to test more complicated adjustable profiles, forward modeling
quickly becomes too computationally expensive as the dimensionality of the
search space increases. We need to devise a more efficient approach to
explore the myriad possibilities.

% FIGURE 5.1 %%%%%%%%%%%%%%%%%%%%%%%%%%%%%%%%%%%%%%%%%%%%%%%%%%%%%%%%%%%%%%%%
\begin{figure}[b]
\hskip 0.625in
\epsfxsize 4.25in
\epsffile{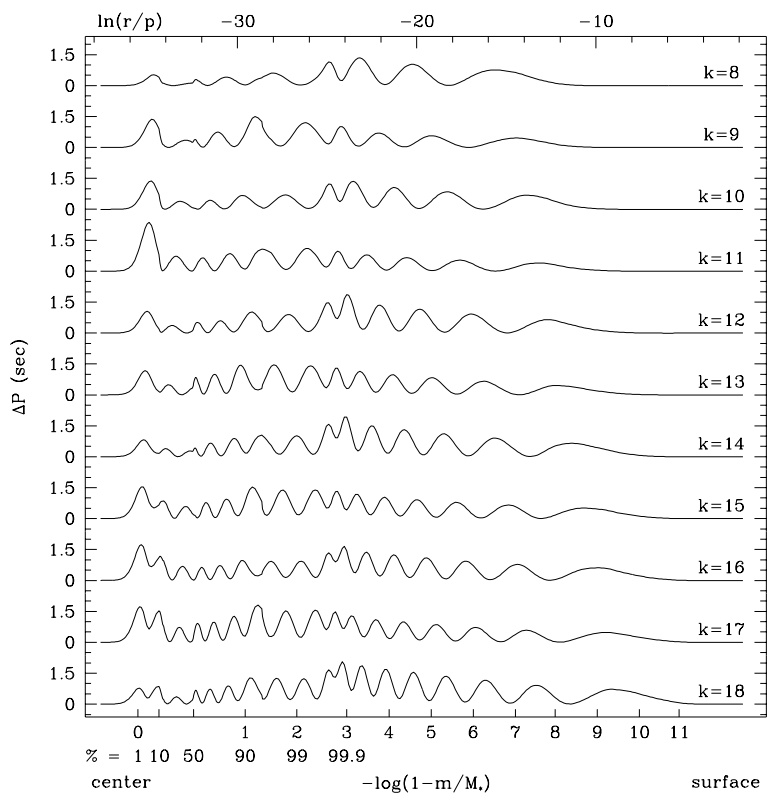}
\caption[Sensitivity of pulsation periods to the deep internal structure]
{For the best-fit model of GD~358 from Table \ref{tab4.2} this plot shows
the change in pulsation period for $\ell=1$ modes of various radial
overtone number ($k$) which result from a smooth artificial 10\% decrease
in the Brunt-V\"ais\"al\"a frequency as a function of the natural log of
the ratio of the distance from the center of the model to the local
pressure (top axis) and the fractional mass coordinate $-\log(1-m/M_*)$
(bottom axis). The center of the model is to the left, and the surface is
to the right. Also indicated is the mass fraction expressed as a
percentage for several values closer to the center.\label{fig5.1}}
\end{figure}
%%%%%%%%%%%%%%%%%%%%%%%%%%%%%%%%%%%%%%%%%%%%%%%%%%%%%%%%%%%%%%%%%%%%%%%%%%%

The natural frequency that dominates the determination of pulsation
periods in white dwarf models is the Brunt-V\"ais\"al\"a (BV) frequency,
which we calculate using the Modified Ledoux treatment described in
\cite{tfw90}. To get a sense of how the pulsation periods depend on the BV
frequency in various regions of the model interior, we added a smooth
perturbation to the best-fit model of GD~358 from Table \ref{tab4.2},
moving it one shell at a time from the center to the surface. The
perturbation artificially decreased the BV frequency across seven shells,
with a maximum amplitude of 10 percent. We monitored the effect on each of
the pulsation periods as the perturbation moved outward through the
interior of the model. The results of this experiment for the pulsation
periods corresponding to those observed in GD~358 are shown in Figure
\ref{fig5.1}. Essentially, this experiment demonstrates that the pulsation
periods are sensitive to the conditions all the way down to the inner few
percent of the model. Since the observational uncertainties on each period
are typically only a few hundredths of a second, even small changes to the
BV frequency in the model interior are significant.

\section{Model Perturbations\label{REVSEC}}

The root-mean-square (r.m.s.) residuals between the observed pulsation
periods in GD~358 and those calculated for the best-fit from forward
modeling are still much larger than the observational noise. This suggests
that either we have left something important out of our model, or we have
neglected to optimize one or more of the parameters that could in
principle yield a closer match to the observations. To investigate the
latter possibility, we introduced {\it ad hoc} perturbations to the BV
frequency of the best-fit model to see if the match could be improved.  
Initially, we concentrated on the region of the Brunt-V\"ais\"al\"a curve
that corresponds to the internal chemical profile.

\begin{figure}
\hskip 0.75in
\epsfxsize 4.0in
\epsffile{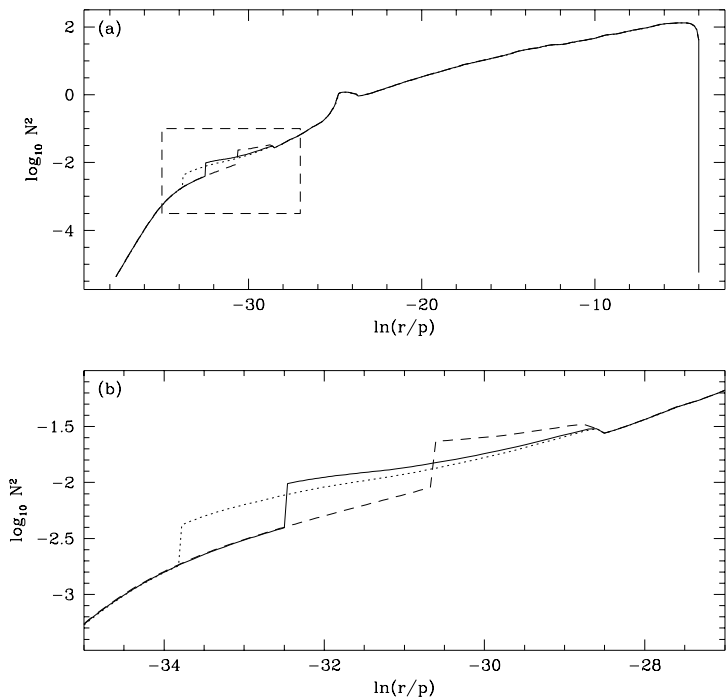}
\caption[The Brunt-V\"ais\"al\"a frequency for various chemical profiles]
{The Brunt-V\"ais\"al\"a frequency as a function of the radial coordinate
ln(r/p) for several models with the same mass, temperature, helium layer
mass, and central oxygen mass fraction but different internal chemical
profiles (a) from the center of the model at left to the surface at right,
and (b) only in the range of ln(r/p) indicated by the dashed box in the
upper panel. The three curves correspond to a profile with $q$ equal to
0.2 (dotted), 0.49 (solid), and 0.8 (dashed).\label{fig5.2}}
\end{figure}

If we look at the BV frequency for models with the same mass, temperature,
helium layer mass, and central oxygen mass fraction but different internal
chemical profiles (see Figure \ref{fig5.2}) it becomes clear that the
differences are localized. In general, we find that changes in the
composition gradient cause shifts in the BV frequency. Moving across an
interface where the gradient becomes steeper, the BV frequency shifts
higher; at an interface where the gradient becomes more shallow, the BV
frequency shifts lower. The greater the change in the gradient, the larger
the shift in the BV frequency.

% FIGURE 5.2 %%%%%%%%%%%%%%%%%%%%%%%%%%%%%%%%%%%%%%%%%%%%%%%%%%%%%%%%%%%%%%%%
%%%%%%%%%%%%%%%%%%%%%%%%%%%%%%%%%%%%%%%%%%%%%%%%%%%%%%%%%%%%%%%%%%%%%%%%%%%

\subsection{Proof of Concept}

We began by generating a model with the same mass, temperature, helium
layer mass, and internal composition as the best-fit from Table
\ref{tab4.2}, but using a uniform internal chemical profile with constant
20:80 C/O out to the 95 percent mass point. We identified a sequence of
100 shells in this model spanning a range of fractional mass from 0.20 to
0.97 and perturbed the BV frequency to try to produce a better match to
the observations. We parameterized the perturbation as a linearly varying
multiplicative factor applied to the BV frequency over a range of shells, 
described by four parameters: (1) the innermost shell to perturb, (2) the 
magnitude of the perturbation at the innermost shell, (3) the number of 
shells in the perturbation range, and (4) the magnitude of the perturbation 
at the outermost shell.  These 4 parameters are sufficient to describe a 
profile with two abrupt changes in the composition gradient.

The innermost shell was allowed to be any of the 100 shells, and the
number of shells in the perturbation range was allowed to be between 0 and
100. If the chosen range would introduce perturbations outside of the
sequence of 100 shells, the outermost of these shells was the last to be
perturbed. The magnitude of the perturbation was allowed to be a
multiplicative factor between 1.0 and 3.0 at both the innermost and
outermost perturbed shells, and was interpolated linearly in the shells
between them.

By using a GA to optimize the perturbation, we can try many random
possibilities and eventually find the global best-fit with far fewer model
evaluations than a full grid search of the parameter-space. We demonstrated 
this by introducing a particular perturbation to the model and determining 
the pulsation periods. Using the same unperturbed model, we then attempted 
to find the parameters of the perturbation by matching the periods using 
the GA. In 9 out of 10 runs (500 generations of 64 trials), the GA found a
first-order solution within two grid points of the input perturbation.  
Thus, the GA combined with a small (625 point) grid search yields a 90
percent probability of success for an individual run. By repeating the
process several times, the probability of finding the correct answer
quickly exceeds 99.9 percent, even while the number of model evaluations
required remains hundreds of times lower than a full grid of the
parameter-space.

\subsection{Application to GD~358}

Having demonstrated that the method works on calculated model periods, we
applied it to the observed pulsation periods of GD~358. We began with a
model similar to the 5-parameter best-fit determined from forward
modeling, but again using a uniform internal chemical profile (constant
16:84 C/O out to 0.95 $m/M_*$). After the GA had found the best-fit
perturbation for GD~358, we reverse-engineered the corresponding chemical
profile.

To accomplish this, we first looked in the unperturbed model for the
fractional mass corresponding to the innermost and outermost shells in the
perturbation range. We fixed the oxygen abundance to that of the
unperturbed model from the center out to the fractional mass of the
innermost perturbed shell. The size of the shift in the BV frequency is
determined by how much the composition gradient changes at this point, so
we adjusted the oxygen abundance at the fractional mass of the outermost
perturbed shell until the change in the gradient produced the required
shift. Finally, we fixed the oxygen abundance to that value from the
outermost perturbed shell out to a fractional mass of 0.95, where it
abruptly goes to zero.

% TABLE 5.1 %%%%%%%%%%%%%%%%%%%%%%%%%%%%%%%%%%%%%%%%%%%%%%%%%%%%%%%%%%%%%%%
\begin{table}[b]
\begin{center}
\caption{Periods and Period Spacings for GD~358 and Best-fit Models
\label{tab5.1}}
\vskip 5pt
\begin{tabular}{lllccccccccc}
\hline\hline
&\multicolumn{2}{c}{Observed} &&\multicolumn{2}{c}{3-par fit} &
&\multicolumn{2}{c}{5-par fit}&&\multicolumn{2}{c}{7-par fit}     \\ 
\cline{2-3}\cline{5-6}\cline{8-9}\cline{11-12}
$k$ &$P$&$\Delta P$& &$P$&$\Delta P$& &$P$&$\Delta P$& &$P$&$\Delta P$  \\
\hline
08$\ldots$ & 423.27 & 40.96 && 422.31 & 42.26 && 422.36 & 40.92 && 422.75 & 39.69 \\
09$\ldots$ & 464.23 & 37.36 && 464.57 & 36.77 && 463.28 & 38.01 && 462.43 & 37.46 \\
10$\ldots$ & 501.59 & 40.16 && 501.35 & 35.88 && 501.29 & 37.50 && 499.90 & 39.70 \\
11$\ldots$ & 541.75 & 35.01 && 537.23 & 39.04 && 538.79 & 36.77 && 539.60 & 36.65 \\
12$\ldots$ & 576.76 & 41.52 && 576.27 & 42.79 && 575.56 & 43.33 && 576.25 & 42.10 \\
13$\ldots$ & 618.28 & 40.07 && 619.06 & 39.79 && 618.89 & 39.85 && 618.36 & 40.25 \\
14$\ldots$ & 658.35 & 42.29 && 658.85 & 42.97 && 658.74 & 42.36 && 658.61 & 42.90 \\
15$\ldots$ & 700.64 & 33.66 && 701.82 & 32.76 && 701.10 & 34.33 && 701.51 & 33.44 \\
16$\ldots$ & 734.30 & 36.37 && 734.58 & 36.92 && 735.42 & 36.99 && 734.95 & 36.59 \\
17$\ldots$ & 770.67 & 40.03 && 771.50 & 39.30 && 772.41 & 37.34 && 771.54 & 39.60 \\
18$\ldots$ & 810.7  & 44.1  && 810.80 & 44.34 && 809.75 & 44.63 && 811.14 & 44.15 \\
\hline\hline
\end{tabular}
\end{center}
\end{table}
%%%%%%%%%%%%%%%%%%%%%%%%%%%%%%%%%%%%%%%%%%%%%%%%%%%%%%%%%%%%%%%%%%%%%%%%%%%

After we found the C/O profile of the best-fit perturbation in this way,
we fixed this reverse-engineered profile in the models and performed a new
fit from forward modeling with the GA to re-optimize the mass,
temperature, helium layer mass, and central oxygen mass fraction. The BV
curve of the final model differs slightly, of course, from that of the
original uniform composition model with the perturbation added. But the
approximate internal structure is preserved, and leads to a better match
to the observed pulsation periods than we could have otherwise found.

\vspace*{-12pt}
\section{Results}

\begin{figure}[p]
\hskip 0.25in
\epsfxsize 5.0in
\epsffile{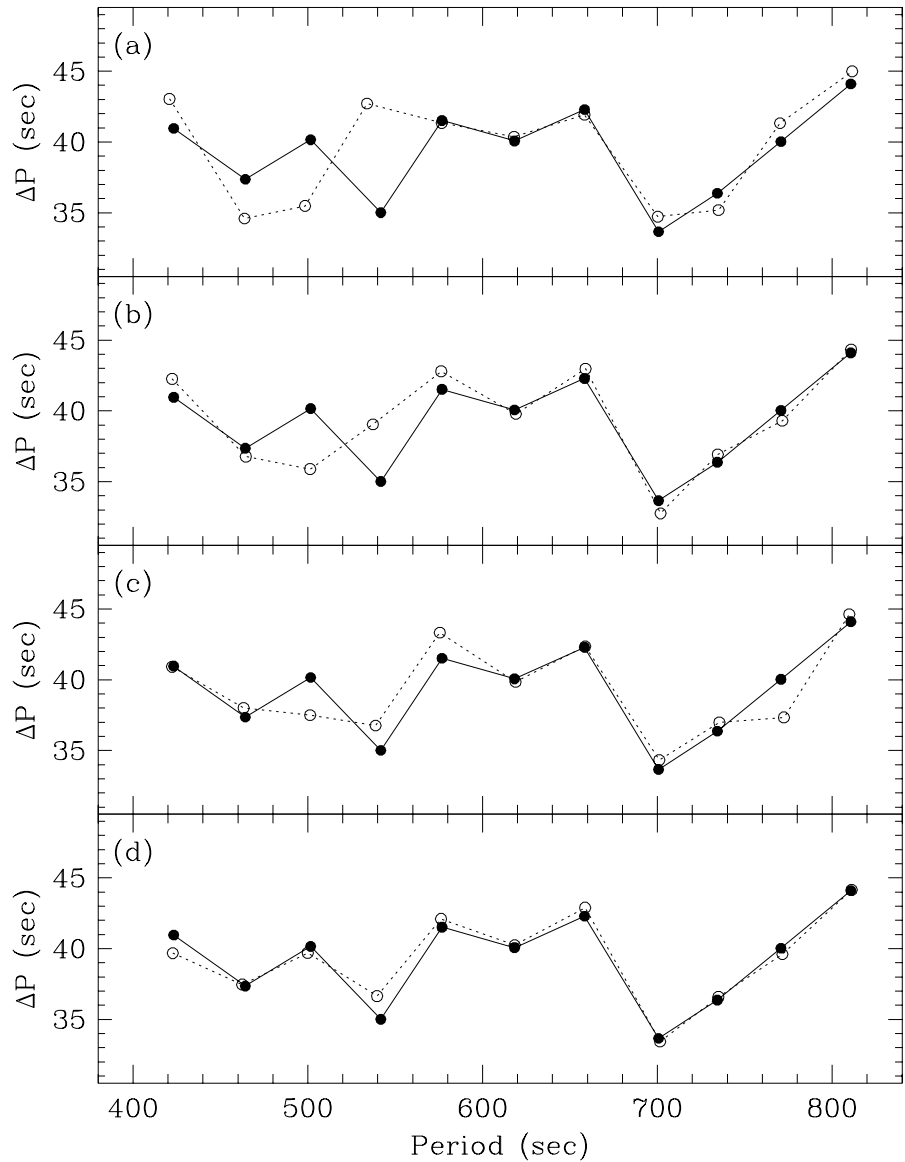}
\caption[GD~358 pulsation periods with various best-fit models]
{The periods and period spacings observed in GD~358 (solid points) with
the theoretical best-fit models (open points) from (a) \cite{bw94b}, (b)
\cite{mnw00}, (c) the 5-parameter forward modeling presented in \S
\ref{FWDSEC}, and (d) the reverse approach. Uncertainties on the
observations are smaller than the size of the points in this
figure.\label{fig5.3}}
\end{figure}

The calculated periods and period spacings ($\Delta P = P_{k+1} - P_k$)  
for the best-fit models from the 5-parameter forward modeling and from the
reverse approach are shown in the bottom two panels of Figure \ref{fig5.3}
along with the data for GD~358. The best-fit models of \cite{bw94b} and
\cite{mnw00} are shown in the top two panels for comparison. The data in 
Figure \ref{fig5.3} for the observations and our best-fit models are given 
in Table \ref{tab5.1}. Some of the improvement evident in the panels of 
Figure \ref{fig5.3} is certainly due to the fact that we have increased 
the number of free parameters. To evaluate whether or not the new models 
represent a {\it significant} improvement to the fit, we use the Bayes 
Information Criterion (BIC), following \cite{mmw01}.

% FIGURE 5.3 %%%%%%%%%%%%%%%%%%%%%%%%%%%%%%%%%%%%%%%%%%%%%%%%%%%%%%%%%%%%%%
%%%%%%%%%%%%%%%%%%%%%%%%%%%%%%%%%%%%%%%%%%%%%%%%%%%%%%%%%%%%%%%%%%%%%%%%%%%

The fits listed in Table \ref{tab4.2} used $n_p=3$ completely free
parameters, sampling several combinations of two additional parameters.  
This amounts to a partial optimization in 5 dimensions. To make a fair
comparison we use the best carbon-core model from Table \ref{tab4.2},
which represents the best truly 3-parameter fit. This model had r.m.s.
residuals of $\sigma(P)=2.30$ seconds for the periods and $\sigma(\Delta
P)=2.65$ seconds for the period spacings. For $N=11$ data points, the BIC
leads us to expect the residuals of a $n_p=5$ fit to decrease to
$\sigma(P)=1.84$ and $\sigma(\Delta P)=2.13$ just from the addition of the
extra parameters. In fact, the fit from the forward modeling presented in
\S \ref{FWDSEC} has $\sigma(P)=1.28$ and $\sigma(\Delta P)=1.42$, so we
conclude that the improvement is statistically significant.

The results of the reverse approach presented in \S \ref{REVSEC} are
harder to evaluate because we are perturbing the BV frequency directly,
rather than through a specific parameter. We consider each additional
point in the internal chemical profile where the composition gradient
changes to be a free parameter. Under this definition, the perturbed
models are equivalent to a 7-parameter fit since there are three such
points in the profiles, compared to only one for the 5-parameter case. If
we again use the BIC, we expect the residuals to decrease from their
$n_p=5$ values to $\sigma(P)=1.03$ and $\sigma(\Delta P)=1.14$ seconds.  
After re-optimizing the other four parameters using the profile inferred
from the reverse approach, the residuals actually decreased to
$\sigma(P)=1.11$ and $\sigma(\Delta P)=0.71$ seconds. The decrease in the
period residuals is not significant, but the period spacings are improved
considerably. This is evident in the bottom panel of Figure \ref{fig5.3}.

\section{Chemical Profiles}

\begin{figure}[p]
\epsfxsize 5.7in
\epsffile{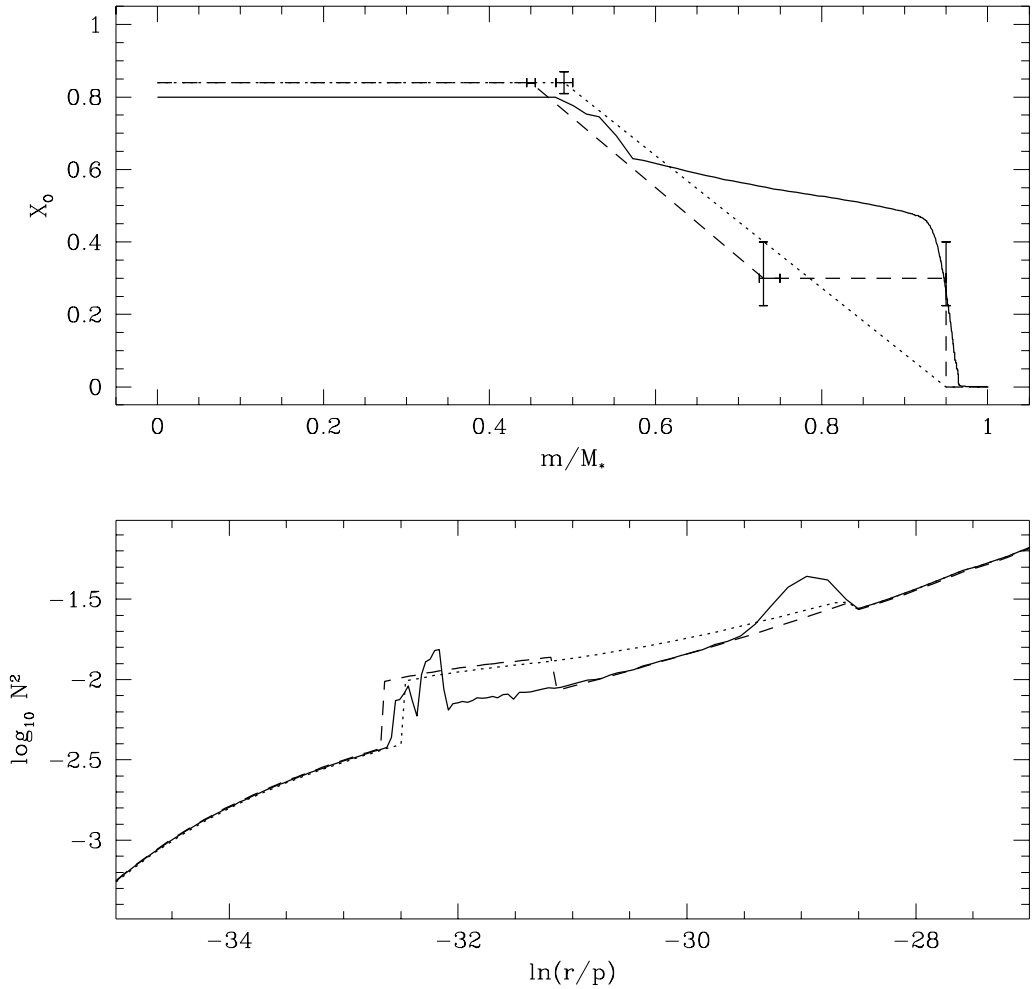}
\caption[Internal chemical profiles of the best-fit models to GD~358]
{The internal oxygen profiles (top) and the 
corresponding region of the Brunt-V\"ais\"al\"a curves (bottom) for the 
best-fit forward model from \S \ref{FWDSEC} (dotted), the result of the
best-fit reverse approach from \S \ref{REVSEC} (dashed), and the scaled
theoretical calculations of \cite{sal97} for comparison (solid).
\label{fig5.4}}
\end{figure}

The internal chemical profiles corresponding to the best-fit models from
the 5-parameter forward modeling and from the reverse approach are shown
in Figure \ref{fig5.4} with the theoretical profile for a 0.61 $M_{\odot}$
model from \cite{sal97}, scaled to a central oxygen mass fraction of 0.80.
The profile from the best-fit forward modeling matches the location and
slope of the initial shallow decrease in the theoretical profile. The
reverse approach also finds significant structure in this region of the
model, and is qualitatively similar to the \cite{sal97} profile to the
extent that our parameterization allows. It is encouraging that both
approaches agree with each other and bear some resemblance to the models.

% FIGURE 5.4 %%%%%%%%%%%%%%%%%%%%%%%%%%%%%%%%%%%%%%%%%%%%%%%%%%%%%%%%%%%%%%
%%%%%%%%%%%%%%%%%%%%%%%%%%%%%%%%%%%%%%%%%%%%%%%%%%%%%%%%%%%%%%%%%%%%%%%%%%%

\vfill\cleardoublepage\thispagestyle{empty}

\chapter{Conclusions \label{conc}}

\begin{quote}
``The reasonable man adapts himself to the world; the unreasonable one 
persists in trying to adapt the world to himself.  Therefore all progress 
depends on the unreasonable man.'' 

\hskip 2.5in{\it ---George Bernard \citeauthor{sha03}}
\end{quote}

\section{Discussion of Results}

The application of genetic-algorithm-based optimization to white dwarf
pulsation models has turned out to be very fruitful. We are now confident
that we can rely on this approach to perform global searches and to
provide not only objective, global best-fit models for the observed
pulsation frequencies of DBV white dwarfs, but also fairly detailed maps
of the parameter-space as a natural byproduct. This approach can easily be
extended to treat the DAV stars and, with a grid of more detailed starter
models, eventually the DOVs. Ongoing all-sky surveys promise to yield many
new pulsating white dwarfs of all classes which will require follow-up
with the Whole Earth Telescope to obtain seismological data. With the
observation and analysis procedures in place, we will quickly be able to
understand the statistical properties of these ubiquitous and relatively
simple stellar objects. Our initial 3-parameter application of the method
provided new evidence that the pulsation frequencies of white dwarfs
really are global oscillations. We refined our knowledge of the
sensitivity of the models to the structure of the envelope, and we
demonstrated that they are sensitive to the conditions deep in the
interior of the star, as suggested by previous work on crystallization by
\cite{mw99}.

The extension of the genetic-algorithm-based approach to optimize the
internal composition and structure of our models of GD~358 yielded more
exciting results. The values of the 3 parameters considered in the initial
study ($M_*, T_{\rm eff}, M_{\rm He}$) were unchanged in the full
5-parameter fit, so we feel confident that they are the most important for
matching the gross period structure. The efficiency of the GA relative to
a grid search was much higher for this larger parameter-space, and the
ability of the method to find the global best-fit was undiminished. The
significant improvement to the fit made possible by including $X_{\rm O}$
and $q$ as free parameters confirms that the observed pulsations really do
contain information about the hidden interiors of these stars.

Our best-fit solution has a thick helium layer, which should help to
resolve the controversy surrounding the evolutionary connection between
the PG~1159 stars and the DBVs. The helium layer mass for PG~1159-035 from
the asteroseismological investigation of \cite{win91} was relatively
thick, at $\sim3\times10^{-3}$~\Msun. \cite{kle98} found good agreement
with the observed pulsations in the DAV star G29--38 using a similar
helium layer mass. If the standard picture of white dwarf evolution is
correct, with a slow cooling process connecting all three classes of
pulsators, then we would expect a similar structure for the DBVs. The
original best-fit model for GD~358 by \cite{bw94b} had a relatively thin
helium layer, at $\sim1.2\times10^{-6}$~\Msun. This posed a problem for
the standard picture. \cite{dk95} treated this problem by including
time-dependent diffusive processes in their calculations, but admitted
that it still could not explain the presence of the DB gap, which is still
an unresolved problem.  Our thick envelope solution also fits more
comfortably within the evolutionary scenario of a hot DB star becoming a
carbon (DQ) white dwarf without developing an anomalously high
photospheric carbon abundance \citep{pro00}.

We have finally measured the central oxygen abundance in GD~358 and used
it to provide a preliminary constraint on the $^{12}$C$(\alpha ,\gamma
)^{16}$O nuclear reaction cross-section. This reaction is one of the most
important for understanding the late stages of stellar evolution and
supernovae. Our preliminary value for the astrophysical S-factor at 300
keV ($S_{300}=290 \pm 15$ keV barns) is high relative to most published
values. However, recent work on type Ia supernovae also favors a high
value to produce model light curves with a sufficiently slow rise to
maximum light \citep{hwt98}. Fortunately, there are other observational
consequences of the higher rate in the spectra of type Ia supernovae
models, so independent evidence should soon be possible.

More precise constraints on $^{12}$C$(\alpha,\gamma)^{16}$O from
asteroseismology will require additional detailed simulations like those
of \cite{sal97}. By determining the range of values for the cross-section
that produce a central oxygen abundance within the measurement
uncertainties of $X_{\rm O}$, we should be able to surpass the precision
of the extrapolation from laboratory measurements by nearly an order of
magnitude. The quoted uncertainty on our preliminary measurement of the
$^{12}$C$(\alpha,\gamma)^{16}$O cross-section does not include systematic
effects. There will certainly be some error associated with using our
white dwarf models; we already know that they aren't perfect.  There will
also be some contribution to the uncertainty from the assumptions built in
to the chemical profiles of \cite{sal97}, particularly from the
description of convection.

We have demonstrated that the pulsation periods in our white dwarf models
are sensitive to the shape of the internal chemical profiles. We can use
this shape as a powerful diagnostic of other physical processes relevant
to white dwarf model interiors, such as convective overshooting and
crystallization.

While they are still embedded in the cores of red giant models, the
internal chemical profiles of white dwarf models show a relatively
constant C/O ratio near the center, with a size determined by the extent
of the helium-burning convective region. The degree of mixing at the edge
of this region is unknown, so a convective overshooting parameter is used
to investigate the effect of different assumptions about mixing. With no
convective overshooting, the final C/O ratio is constant out to the 50\%
mass point; with the convective overshooting parameter fixed at an
empirically derived value, the central oxygen mass fraction is unchanged a
the level of a few percent, but the region with a constant C/O ratio
extends out to the 65\% mass point. Further out in both cases the oxygen
mass fraction decreases as a result of the helium-burning shell moving
toward the surface of the red giant model while gravitational contraction
causes the temperature and density to rise. This increases the efficiency
of the triple-$\alpha$ reaction, producing more carbon relative to oxygen.

Our parameterization of the internal chemical profile is not yet detailed
enough to probe all of the physical information contained in the actual
profiles. Our results suggest that convective overshooting is not required
to explain the internal chemical profile of GD~358, to the extent that we
can measure it at this time. Additional fitting with more detailed
evolutionary profiles will provide a definite statement about convective
overshooting, and will also provide constraints on the
$^{12}$C$(\alpha,\gamma)^{16}$O reaction over the range of temperatures
and densities sampled during the helium shell-burning phase.

Measurements of the internal chemical profiles will also provide a test of
phase separation and crystallization in more massive or cooler pulsating
white dwarf stars. The distribution of oxygen in the interior of a
crystallizing white dwarf model is significantly different from the
chemical profile during the liquid phase. The central oxygen mass fraction
is higher, and the structure in the profile virtually disappears
\citep{sal97}.

Our constraints presently come from measurements of a single white dwarf
star. Application of the GA fitting method to additional pulsating white
dwarfs will provide independent determinations of the central C/O ratio
and internal chemical profiles. These measurements should lead to the same
nuclear physics, or something is seriously wrong. Either way, we will
learn something useful. It would be best to apply this technique to
another DBV star before applying it to another class of pulsators, since
it is still not certain that all of them are produced in the same way. If
we were to find a significantly different C/O ratio for another kind of 
pulsator, it could be telling us something about differences in the 
formation mechanisms.

The reverse approach to model-fitting has opened the door to exploring
more complicated chemical profiles, and the initial results show
qualitative agreement with recent theoretical calculations. We were
originally motivated to develop this approach because the variance of our
best-fit model from the forward method was still far larger than the
observational uncertainty. This initial application has demonstrated the
clear potential of the approach to yield better fits to the data, but the
improvement to the residuals was only marginally significant. We should
continue to develop this technique, but we must simultaneously work to
improve the input physics of our models. In particular, we should really
include oxygen in our envelopes and eventually calculate fully
self-consistent models out to the surface.

\section{The Future}

\subsection{Next Generation Metacomputers}

In the short time since we finished building the specialized parallel
computer that made the genetic algorithm approach feasible, processor and
bus speeds have both quadrupled. At the same time, multiple-processor main
boards have become significantly less expensive and operating systems have
enhanced their shared memory multi-processing capabilities. These
developments imply that a new metacomputer with only 16 processors on as
few as 4 boards could now yield an equivalent computing power in a smaller
space at a reduced price. There's no shame in this, it's just the nature
of computer technology.

A famous empirical relation known as Moore's Law notes that computing
speed doubles every 18 months. This has been true since the 1960's. A
group at Steward Observatory recently used this relation to determine the
largest calculation that should be attempted at any given time
\citep{got99}. Their premise was that since computing power is always
growing, it is sometimes more efficient to wait for technology to improve
before beginning a calculation, rather than using the current technology.
They found that any computation requiring longer than 26 months should not
be attempted using presently available technology. We are happy to report
that our calculations fell below this threshold, so we are better off
today than we would have been if we had spent a year at the beach before
building the metacomputer to do this project.

The guiding principle we used three years ago attempted to maximize the
computing power of the machine per dollar. We now believe there are
additional factors that should be considered. First, the marginal cost of
buying presently available technology that will allow for easy upgrades in
the future (especially faster processors) is small relative to the
replacement cost of outdated parts. For a few hundred dollars extra, we
could have bought 100 MHz motherboards that would have allowed us to
nearly triple the speed of the machine today for only the cost of the
processors.

Second, the marginal cost of buying quality hardware (especially power
supplies and memory) rather than the absolute cheapest hardware available
is small relative to the cost of time spent replacing the cheap hardware.
We actually learned this lesson before buying the hardware for the top row
of 32 nodes. We have never had to replace a power supply in one of the top
32, but the bottom 32 still have power supply failures at the rate of one
every two months or so.

Finally, because some hardware problems are inevitable, it pays to keep
the number of nodes as small as possible. The marginal cost of slightly
faster processors may be small compared to the time spent fixing problems
on a larger number of nodes. True, when one of many nodes runs into a
problem it has a smaller effect on the total computing power available,
but the frequency of such problems is also higher. If we were to build
another metacomputer today, we would estimate not only our budget in
dollars, but also our budget in time.

\subsection{Code Adaptations}

Initially, we did not believe it would be possible to run the parallel
genetic algorithm on supercomputers because, in its current form, the code
dynamically spawns new tasks throughout the run. On local supercomputers
at the University of Texas, software restrictions in the Cray
implementation of PVM allow only a fixed number of slave tasks to be
spawned, and only at the beginning of a run. This feature is intended to
prevent any single job from dominating the resources.

Since the metacomputer provided a more cost effective solution to our
computing requirements at the time, we never revisited the problem. We now
believe that relatively minor modifications to the code would allow the
slave jobs to remain active on a fixed number of processing elements and
retain their essential function. This could allow us to solve much larger
problems in a shorter time if we have access to supercomputers in the
future.

Eventually, we hope to develop a more advanced version of the PIKAIA
genetic algorithm. In particular, we'd like to incorporate a hybrid
routine that would use a local hill-climbing scheme to speed up the
convergence after the initial success of the genetic operators.

\subsection{More Forward Modeling}

In addition to the immediate application of forward modeling to more DBV
white dwarfs, there are several possible extensions of the method to other
types of white dwarf stars.

To use white dwarfs effectively as independent chronometers for stellar
populations, we need to calibrate evolution models with observations of
the internal structure of the coolest white dwarfs. The
hydrogen-atmosphere variable (DAV) white dwarfs are the coolest class of
known pulsators, so they can provide the most stringent constraints on the
models.

Previous attempts to understand these objects have been hampered by their
relatively sparse pulsation spectra. \cite{kle98} made repeated
observations of the star G29--38 over many years and found a stable
underlying frequency structure, even though only a subset of the full
spectrum of modes were visible in each data set. Preliminary attempts to
match the complete set of frequencies have focused on calculating grids of
DAV models, but the huge range of possible parameters makes this task very
computationally intensive. We hope to use the genetic-algorithm-based
approach to explore the problem more efficiently.

\cite{mon99} showed that phase separation in crystallizing white dwarfs
could add as much as 1.5 Gyr to age estimates. With the discovery of the
massive pulsator BPM 37093 by \cite{kan92}, we now have the opportunity to
test the theory of crystallization directly and calibrate this major
uncertainty.

Like most other hydrogen-atmosphere white dwarfs, BPM 37093 exhibits a
limited number of excited pulsation modes, but \cite{nit00} secured
reliable identifications of the spherical degree of these modes using the
Hubble Space Telescope during a ground-based WET campaign. Preliminary
attempts by \cite{kan00} to match the frequency structure revealed a
parameter-correlation between the crystallized mass fraction and the
thickness of the hydrogen layer.

The genetic algorithm is well equipped to deal with parameter-correlations.  
The initial application to GD~358 revealed correlations between several 
parameters and helped us to understand them in terms of the basic physical 
properties of the model. Despite the correlations, the genetic algorithm 
consistently found the global solution in every test using synthetic data, 
so we are confident that we will be able to use this method to separate 
unambiguously the effects of stellar crystallization from other model 
parameters.

\subsection{Ultimate Limits of Asteroseismology}

In our initial application of the reverse approach, we concentrated on
only one region of the Brunt-V\"ais\"al\"a curve and we parameterized the
perturbation to explore different possible internal chemical profiles
efficiently. The technique is clearly useful, and we hope to use it to
investigate a broad range of characteristics in our models that would be
impractical to approach through forward modeling. In particular, we hope
to quantify the ultimate limits of asteroseismology---to determine what we
can learn from the data, and what we can never learn.

By using perturbations with various parameterizations, we may be able to
probe weaknesses in the models themselves. We can address the question of
what limitations we are imposing on our understanding simply by using the
models we use. We may find that a whole range of models are pulsationally
indistinguishable, or perhaps that our knowledge of certain regions of the
model interior are limited only by the particular pulsation modes that we
observe in the stars. It will be an entirely new way of looking at the
problem, and it will give us the opportunity to learn even more from our
data.

\section{Overview}

At the beginning of this project, we set out to learn something about
nuclear fusion using pulsating white dwarf stars as a laboratory. What
we've learned is unlikely to allow humanity to manufacture clean
sustainable energy anytime soon; but the project has demonstrated that
white dwarf asteroseismology has a clear potential to improve our
knowledge of some fundamental nuclear physics. This is definitely a step
in the right direction.

Along the way, we've developed some tools that hold great promise for the
future of white dwarf asteroseismology and other computationally intensive
modeling applications. We have developed a minimal-hardware design for a
scalable parallel computer based on inexpensive off-the-shelf components.
We have documented the server-side software configuration required to
operate such a machine, and we have developed a generalized parallel
genetic algorithm which can exploit the full potential of the hardware. We
have modified well-established white dwarf evolution and pulsation codes
to interface with the parallel GA and to provide stable operation over a
broad range of interesting physical parameters.

We have laid the groundwork for new analyses that promise to unlock the
secrets of the white dwarf stars. The fun has only started.

\vfill\cleardoublepage\thispagestyle{empty}

\appendix

\chapter{Observations for the WET \label{wet}}

\newpage

\begin{figure}[t]
\hskip 0.1in
\epsfxsize 5.5in
\epsffile{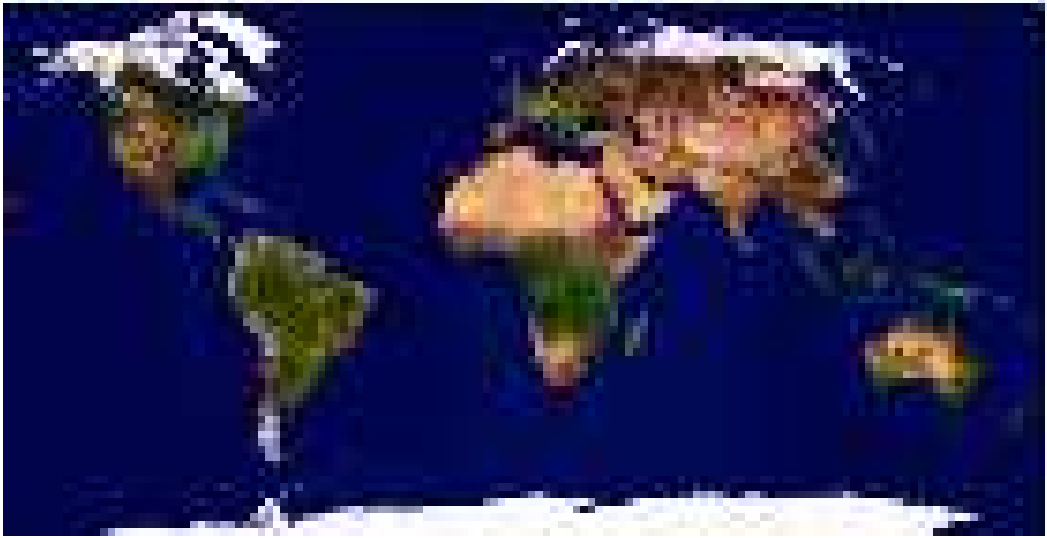}
\caption[The Whole Earth Telescope]
{The Whole Earth Telescope.\label{figa.1}}
\end{figure}

\vspace*{0.4in}

\section{What is the WET?}

The Whole Earth Telescope (WET) is an informal collaboration of
astronomers at observatories around the world who cooperate to produce
nearly continuous time-series photometry of white dwarfs and similar
targets for up to 14 days at a time (see Figure \ref{figa.1}).  This
instrument has been operating since 1988, and is currently run from
headquarters at Iowa State University.

During my time as a graduate student, I have participated in four
organized campaigns to observe white dwarfs with the WET. Each campaign is
referred to as XCOV (for extended coverage) followed by a number. XCOV 1
took place in March 1988. In every case but one, I was stationed at the
2.1-meter telescope at McDonald Observatory in west Texas. For XCOV 17, I
used the 1.5-meter telescope at Cerro Tololo Interamerican Observatory.
For the electronic edition of my dissertation, I have archived all of the
raw observations that I obtained for each of these campaigns.

\newpage

\vspace*{0.5in}

\section{XCOV 15: DQ Herculis}

archive in digital dissertation:[tsm-0023 $\rightarrow$ tsm-0032]

\vspace*{0.5in}

\section{XCOV 17: BPM 37093}

archive in digital dissertation:[tsm-0033 $\rightarrow$ tsm-0048]

\vspace*{0.5in}

\section{XCOV 18: HL Tau 76}

archive in digital dissertation:[tsm-0049 $\rightarrow$ tsm-0072]

\vspace*{0.5in}

\section{XCOV 19: GD 358}

archive in digital dissertation:[tsm-0074 $\rightarrow$ tsm-0085]

\vfill\cleardoublepage\thispagestyle{empty}

\chapter{Interactive Simulations \label{sim}}

\newpage

\vspace*{1.0in}

\section{Pulsation Visualizations}
In digital dissertation: Spherical harmonic visualizations.

\vspace*{1.0in}

\section{The Effect of Viewing Angle}
In digital dissertation: Interactive demonstration of the effect 
of viewing angle on the amplitude of the observed brightness variations.

\vfill\cleardoublepage\thispagestyle{empty}

\chapter{Computer Codes \label{code}}

\newpage

\noindent 
This appendix contains an archive of the source code for the software 
that I have used for the calculations presented in this dissertation.

EVOLVE.F is a streamlined version of the White Dwarf Evolution Code
(WDEC) described in \S 4.3.2 with references to its origins and to
the sources of data for the input physics. WDEC takes as input a
hot starter model with a specific mass, which can come from detailed 
evolutionary calculations in the case of DOV stars, or from a simple 
polytropic approximation in the case of DBV and DAV stars. Using this 
starter model and other parameters specified in the header, WDEC adds 
an envelope with the specified composition and fractional mass and
evolves the model quasi-statically until it reaches the specified 
temperature.

PULSATE.F uses the final output model produced by WDEC and calculates
the $m=0$ adiabatic non-radial oscillation periods of a specified 
spherical degree ($\ell$) within a specified period range. The periods
resulting from the adiabatic approximation typically differ from the
non-adiabatic results by only a few thousandths of a second, which is
well below the present level of observational noise.

PVM\_FITNESS.F is the code that uses the message-passing routines of
the Parallel Virtual Machine (PVM) software to allow the public-domain
genetic algorithm PIKAIA to evaluate the fitnesses of trials in parallel
rather than sequentially. This code automatically determines the number
of processors available for the calculation, balances the load when 
machines with differing speeds are used, and works around crashed jobs
in a sensible way.

FF\_SLAVE.F is an interface between the parallel genetic algorithm and the 
streamlined version of WDEC. This code runs on each machine that is used
to calculate white dwarf models. It uses the message-passing routines of
PVM to receive sets of parameters from the master process, evaluates the
white dwarf model specified by those parameters, compares the model 
periods to the observations, and returns a measure of fitness to PIKAIA.

The practical aspects of running the evolution and pulsation codes are
addressed in the documentation archive at the end of this appendix.

\newpage

\vspace*{1.0in}

\section{EVOLVE.F}

In digital dissertation: Hypertext version of evolution code.

\vspace*{1.0in}

\section{PULSATE.F}

In digital dissertation: Hypertext version of pulsation code.

\newpage

\section{PVM\_FITNESS.F}

{\small
\begin{verbatim}
      subroutine pvm_fitness (slave, num_jobs, npar, oldph, fitness)
c ---------------------------------------------
c     parallel fitness evaluation using PVM
c ---------------------------------------------
      implicit none
c
      include '../include/fpvm3.h'
c
      integer job, info, nhost, msgtype, iwhich, i
      integer mytid, dtid, tids(0:128), flag, ntask
      integer ttids(64), ptids(64), htids(64), flags(64)
      integer speed, narch, numt, npar, nspawn, last, wait
      integer num_jobs, ndone, length, par, trial, listen
      integer finished(1024),resubmitted(1024)
c
      double precision result, data(64)
      real fitness(1024), oldph(64,1024)
c
      character*40 hostname
      character*18 host
      character*8 slave, arch
      character*8 aout(64)

c ---------------------------------------------
c     initialize book-keeping variables
c ---------------------------------------------
      listen = 0
      wait = 0
      ndone = 0
      do job=1,num_jobs
         finished(job) = 0
         resubmitted(job) = 0
      enddo
c ---------------------------------------------
c     enroll this program in PVM
c ---------------------------------------------
      call pvmfmytid( mytid )
      call pvmfconfig( nhost, narch, dtid, host, arch, speed, info )
c ---------------------------------------------
c     run jobs on slave nodes only
c ---------------------------------------------
      arch = '.'
      flag = PvmTaskHost+PvmHostCompl
      nspawn = nhost-1
      call pvmfspawn( slave, flag, arch, nspawn, tids, numt )
c ---------------------------------------------
c     check for problems spawning slaves
c ---------------------------------------------
      if( numt .lt. nspawn ) then
         write(*,*) 'trouble spawning ',slave
         write(*,*) ' Check tids for error code'
         call shutdown( numt, tids )
      endif
c
      write(*,*)
c ---------------------------------------------
c     send an initial job to each node
c ---------------------------------------------
      do job=0,nspawn-1
c
         trial = job + 1
         do par=1,npar
            data(par) = INT((100*oldph(par,trial))+0.5)/100.
         enddo
c
         call pvmfinitsend( PVMDEFAULT, info )
         call pvmfpack( INTEGER4, trial, 1, 1, info )
         call pvmfpack( INTEGER4, npar, 1, 1, info )
         call pvmfpack( REAL8, data, npar, 1, info ) 
         msgtype  = 1 
         call pvmfsend( tids(job), msgtype, info )
c
 11      format("job ",i3,3(2x,f4.2))
         write(*,11) trial,data(1),data(2),data(3)
c
      enddo
c
      write(*,*)
c
      do job=1,num_jobs
c ---------------------------------------------
c     listen for responses
c ---------------------------------------------
 25      msgtype  = 2 
         call pvmfnrecv( -1, msgtype, info )
         listen = listen + 1
c
         if (info .GT. 0) then
            write(*,*) "<-- job ",job
            listen = 0
            wait = 0
c ---------------------------------------------
c     get data from responding node
c ---------------------------------------------
            call pvmfunpack( INTEGER4, trial, 1, 1, info )
            call pvmfunpack( REAL8, result, 1, 1, info )
            call pvmfunpack( INTEGER4, length, 1, 1, info )
            call pvmfunpack( STRING, hostname, length, 1, info )
c ---------------------------------------------
c     re-send jobs that return crash signal
c ---------------------------------------------
            if ((result .eq. 0.0).and.(resubmitted(trial).ne.1)) then
               write(*,*) "detected fitness=0 job: trial ",trial
               call sendjob
     &         (trial,hostname,'ffrslave',npar,resubmitted,oldph)
               goto 25
            endif
c
            fitness(trial) = result
            finished(trial) = 1
            ndone = ndone + 1
c
 33         format(i4,2x,i4,2x,a8,2x,3(f4.2,2x),f12.8)
            write(*,33) ndone,trial,hostname,oldph(1,trial),
     &                   oldph(2,trial),oldph(3,trial),result
c ---------------------------------------------
c     send new job to responding node
c ---------------------------------------------
 140        if (ndone .LE. (num_jobs-nspawn)) then
               trial = job + nspawn
               call sendjob
     &         (trial,hostname,slave,npar,resubmitted,oldph)
            endif
            goto 100
         endif
c ---------------------------------------------
c     re-submit crashed jobs to free nodes
c ---------------------------------------------
         if (ndone .GT.(num_jobs-nspawn)) then
            last = ndone-nspawn
            if (ndone .GE.(num_jobs-5)) last=ndone
            do trial=1,last
               if ((finished(trial).NE.1).AND.
     &         (resubmitted(trial).NE.1).AND.(wait.NE.1)) then
                  write(*,*) "detected crashed job: trial ",trial
                  call sendjob
     &            (trial,hostname,'ffrslave',npar,resubmitted,oldph)
                  wait = 1
                  goto 25
               endif
            enddo
         endif
c ---------------------------------------------
c     return to listen again or move on
c ---------------------------------------------
         if ((info .EQ. 0).AND.(listen .LT. 10000000)) goto 25
c
         write(*,*) "detected unstable jobs: setting fitness=0"
         do trial=1,num_jobs
            if ((finished(trial) .NE. 1).AND.
     &          (resubmitted(trial) .EQ. 1)) then
               fitness(trial) = 0.0
               finished(trial) = 1
               ndone = ndone + 1
               write(*,33) ndone,trial,hostname,oldph(1,trial),
     &          oldph(2,trial),oldph(3,trial),fitness(trial)
            endif
         enddo
         goto 199
 100     continue
      enddo
c ---------------------------------------------
c     kill any remaining jobs
c ---------------------------------------------
 199  iwhich = PVMDEFAULT
      call pvmftasks( iwhich, ntask, ttids(1), ptids(1),
     &                htids(1), flags(1), aout(1), info )
      do i=2,ntask
         call pvmftasks( iwhich, ntask, ttids(i), ptids(i),
     &                   htids(i), flags(i), aout(i), info )
         if ((aout(i) .EQ. 'ff_slave').OR.
     &       (aout(i) .EQ. 'ffrslave')) then
            call pvmfkill (ttids(i), info)
         endif
      enddo
c
      call pvmfexit(info) 
c
      return
      end
c**********************************************************************
      subroutine sendjob(trial,hostname,slave,npar,resubmitted,oldph)
c
      implicit none
c
      include '../include/fpvm3.h'
c
      integer tids(0:128), numt, msgtype, par, npar, trial, info, flag
      integer resubmitted(1024)
c
      double precision data(64)
      real oldph(64,1024)
c
      character*40 hostname
      character*8 slave
c
      call pvmfspawn( slave, 1, hostname, 1, tids, numt )
c
      if ( numt .lt. 1 ) then
         write(*,*) 'trouble spawning',slave
         write(*,*) ' Check tids for error code'
         call shutdown( numt, tids )
      endif
c
      do par=1,npar
         data(par) = INT((100*oldph(par,trial))+0.5)/100.
      enddo
c
      call pvmfinitsend( PVMDEFAULT, info )
      call pvmfpack( INTEGER4, trial, 1, 1, info )
      call pvmfpack( INTEGER4, npar, 1, 1, info )
      call pvmfpack( REAL8, data, npar, 1, info )
      msgtype  = 1
      call pvmfsend( tids(0), msgtype, info )
c
 55   format("job --> ",a8,3(2x,f4.2))
      write(*,55) hostname,data(1),data(2),data(3)
c
      if (slave .EQ. 'ffrslave') resubmitted(trial) = 1
c
      return
      end
c**********************************************************************
      subroutine shutdown( nproc, tids )
c
      implicit none
c
      integer nproc, i, info, tids(*)
c
      do i=0, nproc
         call pvmfkill( tids(i), info )
      enddo
c
      call pvmfexit( info )
c
      return
      end
c**********************************************************************
\end{verbatim}}

\newpage

\section{FF\_SLAVE.F}

{\small
\begin{verbatim}
      program ff_slave 
c ---------------------------------------------
c fitness function slave program
c ---------------------------------------------
      implicit none
c
      include '../include/fpvm3.h'
c
      integer info, mytid, mtid, msgtype, speed, length, i
      integer n, nhost, narch, dtid, hostid, trial
c
      double precision ff, data(32), result
c
      character*40 hostname,machine,arch
c ---------------------------------------------
c enroll this program in PVM
c ---------------------------------------------
      call pvmfmytid( mytid )
c ---------------------------------------------
c get the master's task id
c ---------------------------------------------
      call pvmfparent( mtid )
c ---------------------------------------------
c receive data from master host
c ---------------------------------------------
      msgtype = 1 
      call pvmfrecv( mtid, msgtype, info ) 
      call pvmfunpack( INTEGER4, trial, 1, 1, info )
      call pvmfunpack( INTEGER4, n, 1, 1, info )
      call pvmfunpack( REAL8, data, n, 1, info ) 
c ---------------------------------------------
c perform calculations with data
c ---------------------------------------------
      result = ff( n, data ) 
c ---------------------------------------------
c send result to master host
c ---------------------------------------------
      call pvmftidtohost( mytid, hostid )
 100  call pvmfconfig( nhost, narch, dtid, hostname, arch, speed, info )
      if (dtid .ne. hostid) goto 100
      length = len(hostname)
      machine = hostname(1:length)
c
      call pvmfinitsend( PVMDEFAULT, info )
      call pvmfpack( INTEGER4, trial, 1, 1, info )
      call pvmfpack( REAL8, result, 1, 1, info )
      call pvmfpack( INTEGER4, length, 1, 1, info )
      call pvmfpack( STRING, machine, length, 1, info )
      msgtype  = 2 
      call pvmfsend( mtid, msgtype, info ) 
c ---------------------------------------------
c leave PVM before exiting
c ---------------------------------------------
      call pvmfexit(info) 
c
      stop
      end
c*********************************************************************
\end{verbatim}}

\newpage

\vspace*{0.5in}

\section{Documentation}

In digital dissertation: Documentation archive.
\begin{itemize}
\item{evolution code: (\underline{PS}/\underline{PDF})}
\item{prep code: (\underline{PS}/\underline{PDF})}
\item{pulsation code: (\underline{PS}/\underline{PDF})}
\end{itemize}

\vfill\cleardoublepage\thispagestyle{empty}

\addcontentsline{toc}{chapter}{Bibliography}
\bibliographystyle{thesis}
\bibliography{thesis}
\vfill\cleardoublepage\thispagestyle{empty}

\addcontentsline{toc}{chapter}{Vita}

\chapter*{Vita}

Travis Scott Metcalfe was born in Seaside, Oregon on 12 October 1973. He 
lived on the central Oregon coast with his parents Jerry and Karen until 
graduating from Newport High School in 1991. In August of that year, he 
moved to Tucson, Arizona to study Astronomy and Physics at the University 
of Arizona, where he received a Bachelor of Science degree in May 1996. 
He came to Austin, Texas in August 1996 to begin graduate studies in the 
Department of Astronomy at the University of Texas where he received a 
Master of Arts degree two years later.

\vspace*{0.5in}
\noindent Permanent Address: 3013 Harris Park Avenue, Austin, Texas 78705

\vspace*{0.5in}
\noindent This dissertation was typeset by the author.

\end{document}